\documentclass[prc, reprint, floatfix, showpacs, nofootinbib, superscriptaddress]{revtex4-2}

\usepackage{amsmath}
\usepackage{graphicx}
\usepackage{multirow}

\newcommand{\sNN}{$\sqrt{s_{\mathrm{NN}}}$~}
\newcommand{\pt}{\ensuremath{\it{p}_{\rm T}}~}
\newcommand{\Et}{\ensuremath{\it{E}_{\rm T}}~}
\newcommand{\kt}{\ensuremath{\it{k}_{\rm T}}~}
\newcommand{\ptJet}{\ensuremath{\it{p}_{\mathrm{T, jet}}}~}
\newcommand{\ptAssoc}{\ensuremath{\it{p}_{\mathrm{T, assoc}}}~}

\newcommand{\ptRecoJet}{\ensuremath{p_{\mathrm{T, jet}}^{\rm RECO}}~}
\newcommand{\ptGenJet}{\ensuremath{p_{\mathrm{T, jet}}^{\rm GEN}}~}
\newcommand{\dphi}{\ensuremath{\Delta\phi}~}
\newcommand{\deta}{\ensuremath{\Delta\eta}~}

\newcommand{\GeV}{GeV/$c$~}

\newcommand{\AuAu}{Au+Au~}

\newcommand{\ns}{near-side~}
\newcommand{\as}{away-side~}

\newcommand{\pttrigrange}[2]{$#1 < p_{\mathrm{T, jet}} < $ #2 GeV/$c$~}
\newcommand{\ptassocrange}[2]{$#1 < p_{\mathrm{T, assoc}} <$ #2 GeV/$c$~}

\newcommand\Tstrut{\rule{0pt}{2.6ex}}         
\newcommand\Bstrut{\rule[-0.9ex]{0pt}{0pt}}   

\begin{document}

\title{Jet-hadron correlations with respect to the event plane in \sNN = 200 GeV \AuAu collisions in STAR}
\collaboration{STAR Collaboration}\noaffiliation

\date{\today}
\begin{abstract}
Angular distributions of charged particles relative to jet axes are studied in \sNN = 200 GeV \AuAu collisions as a function of the jet orientation with respect to the event plane.
This differential study tests the expected path-length dependence of energy loss experienced by a hard-scattered parton as it traverses the hot and dense medium formed in heavy-ion collisions.
A second-order event plane is used in the analysis as an experimental estimate of the reaction plane formed by the collision impact parameter and the beam direction.
Charged-particle jets with $15 < \ptJet <$ 20  and $20 < \ptJet <$ 40 \GeV  were reconstructed with the anti-$\kt$ algorithm with radius parameter setting of $R=0.4$ in the 20-50\% centrality bin to maximize the initial-state eccentricity of the interaction region. The reaction plane fit method is implemented to remove the flow-modulated background with better precision than prior methods. Yields and widths of jet-associated charged-hadron distributions are extracted in three angular bins between the jet axis and the event plane.
The event-plane (EP) dependence is further quantified by ratios of the associated yields in different EP bins. 
No dependence on orientation of the jet axis with respect to the event plane is seen within the uncertainties in the kinematic regime studied. 
This finding is consistent with a similar experimental observation by ALICE in \sNN = 2.76 TeV Pb-Pb collision data.
\end{abstract}
\maketitle
\section{Introduction}

Relativistic heavy-ion collisions have been used for more than three decades to map out the phase diagram of quantum chromodynamics (QCD) matter. This has been done through previous studies from energies around $5-20$ GeV at the Alternating Gradient Synchrotron (AGS) in Brookhaven National Laboratory (BNL) and the Super Proton Synchrotron (SPS) at CERN, to $200$ GeV at the Relativistic Heavy Ion Collider (RHIC) in BNL and upto $5.44$ TeV at the Large Hadron Collider (LHC) at CERN.  A new form of matter has been discovered in such collisions at extreme temperature and density, the ``Quark-Gluon Plasma'' (QGP), that exhibits almost perfect liquid dynamical behavior \cite{Adcox:2004mh,Adams:2005dq,Arsene:2004fa,Back:2004je,Shuryak:2004cy,Lee:2005gw,Gyulassy:2004zy,Heinz:2005zg,Blau:2005pk,Riordan:2006df}. RHIC and the LHC continue to explore new regions of the phase diagram and study the properties of the QGP. 

Observable remnants of partonic interactions at large momentum transfers, called hard probes, travel through the QGP medium and experience energy loss through various QCD interactions with the medium. Hence, they are commonly used to study the structure and dynamics of the QGP~\cite{Gyulassy:1990bh,Gyulassy:2003mc,Wang:1991xy}.  These probes are considered to be highly reliable, due to their expected yields being accurately calculable using the perturbative QCD (pQCD) theoretical framework. Additionally, their short production time ($\tau \sim 1/p_{\mathrm{T}} \leq 0.1$ fm/c) allows for the tracing of medium properties right from the initial phases of the collision. At RHIC, evidence of energy loss in the medium (“jet quenching”) was first observed through properties of leading fragments of jets and their correlations \cite{Adams:2003im,Adams:2006yt,Adler:2006hu}. 

Since 2011, the observation of significant jet quenching has also been confirmed through measurements of reconstructed  back-to-back, inclusive, and tagged jets at the LHC energies \cite{Chatrchyan:2011sx,Chatrchyan:2012gt,CMS:2016uxf,Chatrchyan:2012gw,CMS:2021vui,PhysRevLett.105.252303,JHEP03.2014.13}. The interactions of jets within the hot QCD medium can also be measured experimentally via, for example, the modification of the internal structure of jets, possibly due to medium-induced soft-gluon radiation \cite{Wiedemann:2009sh} and collisional processes\cite{doi:10.1146/annurev.nucl.50.1.37}. The interpretation behind these observations are further supported by correlating jets with charged particles to extend measurements of intrinsic jet properties to large relative angles in \deta and \dphi ~\cite{Khachatryan:2016erx,CMS:2021nhn}. 
More recently, measurements of jet substructure, such as splitting functions that reflect the splitting of a parton into two other partons, and the opening angle of two prongs (where a prong is a jet-like object within a jet), have been studied at LHC and RHIC energies~\cite{Sirunyan_2018,ALargeIonColliderExperiment:2021mqf,ATLASOpenAng,STAR:2021kjt}. The measurements of splitting functions at LHC, for jets with higher transverse momenta, indicate a more unbalanced momentum ratio in central collisions compared to peripheral and $p+p$ collisions. However, at RHIC, the opening angles and splittings of lower momentum jets are found to be vacuum-like, with no quantitative modification in Au+Au collisions compared to reference $p+p$ collisions. The partonic interactions, and therefore medium-induced modifications to a jet, are expected to depend on the path-length traversed by a hard-scattered parton through the medium~\cite{Liou:2013qya}. Leading particles of jets are indeed observed to follow such an expectation, as measured through the azimuthal anisotropy of high transverse momentum ($p_{\rm T}$) hadrons\cite{PhysRevC.76.034904}. However, jet-particle correlations at different angles relative to the event plane at LHC energies have shown no significant path-length dependence of the medium modifications~\cite{PhysRevC.101.064901}. A complimentary study in a lower kinematic range for the jets, accessible at RHIC energies, could provide further constraints on the path-length dependence of jet quenching.

Experimentally, jets are reconstructed by clustering charged-particle tracks and calorimeter-energy depositions using the anti-$\kt$ algorithm\cite{Cacciari:2008gp}. In this analysis, we measure angular correlations of charged-particle tracks with fully reconstructed jets differentially in jet-axis orientation with respect to the reaction plane in \sNN = 200 GeV \AuAu collisions with the STAR experiment.  The reaction plane is defined as the plane formed by the impact parameter and the beam direction.  For non-central collisions of incoming nuclei, the overlap region is an oval ellipsoid, so particles emitted perpendicular to the reaction plane (out-of-plane) have on average a longer length traversed through the medium, than those traveling along the direction of the reaction plane (in-plane).  Studying jets differentially in a relative orientation to the reaction plane allows for a path-length dependent measurement of potential medium modifications.  

In this analysis, the coordinate system used to depict the distribution of associated particles is defined relative to a reconstructed jet, also called a trigger jet. The distribution is thus given by:

\begin{equation}
 \frac{1}{N_{\rm trig}} \frac{d^2 N_{\rm assoc,jet}}{d\dphi d\deta},
 \label{Eq:JHCorrIntro}
\end{equation}

\noindent where $N_{\rm trig}$ is the number of trigger jets, $N_{\rm assoc,jet}$ is the number of associated particles, \dphi ($=|\phi_{\rm jet}-\phi_{\rm assoc}|$) is the azimuthal angle of those associated particles relative to the trigger jets, and \deta ($=|\eta_{\rm jet}-\eta_{\rm assoc}|$) is the difference in the pseudorapidities of the trigger jet and associated particle.  

The goal of this analysis is to study the conditional yield of associated particles, the width of the near- and away-side peaks (quantified using the gaussian width) as a function of the angle between the jet axis and the event plane.  The yield is estimated by:

\begin{equation}
 \mathrm{Yield} = \frac{1}{N_{\rm trig}} \int_{c}^{d} \int_{a}^{b} 
 \frac{d^2N_{\rm assoc,jet}}{d\dphi d\deta}  d\dphi d\deta.
 \label{Eqtn:Yield1}
\end{equation}

\noindent  The choice of integration limits is somewhat arbitrary.  They are chosen based on practical considerations, including the detector acceptance and binning of histograms.  

Selection criteria for events, tracks and towers, along with discussions on track reconstruction efficiency can be found in Sect.~\ref{sect:STARdet}. Measurement of the event plane is discussed in Sect.~\ref{sect:RxnPlane}, followed by details of jet reconstruction in Sect.~\ref{sect:JetReco}. Further details on measuring the correlations between trigger jet and associated hadrons (introduced in Eq.~\ref{Eq:JHCorrIntro}) are given in Sect.~\ref{sect:Correlations}. Background estimation and subtraction done using the Reaction Plane Fit method~\cite{Sharma:2015qra} is discussed in Sect.~\ref{sect:CorrBkgd}. The results are presented in Sect.~\ref{sect:Results}, followed by discussion of the constraints this measurement provides and how it compares to JEWEL~\cite{Zapp:2013vla} calculations and similar measurements at the LHC~\cite{PhysRevC.101.064901}.

\section{Collection of data} \label{sect:STARdet}

A detailed description of the STAR detector and its subsystems can be found in~\cite{Ackermann:2002ad}.  The two sub-detectors used for this analysis, the Time Projection Chamber (TPC) ~\cite{Anderson:2003ur} and the Barrel Electromagnetic Calorimeter (BEMC) ~\cite{BEDDO2003725}, are briefly described in the following.

The TPC detector provides tracking of charged particles over the full azimuthal range with a pseudorapidity coverage of $|\eta| < 1.0$.  Track selection is optimized for track quality and momentum resolution. Reconstructed charged-particle tracks are required to have at least 15 ``hit'' points, and no less than 52\% of the maximum hits possible for a given track kinematics. Tracks are selected as primary if their distance of closest approach (DCA) to the primary vertex is less than 3 cm.  
 Events containing tracks with $\pt >$ 30 \GeV are rejected to avoid contamination from cosmic rays and  mis-reconstruction from fake-tracks.  Tracks with $\pt >$ 2.0 \GeV are used as constituents for jet reconstruction, while tracks with $\pt >$ 1.0 \GeV are used for measuring the correlation functions.  The tracking efficiency is determined from embedding simulations of the detector response and ranges from 75--90\% in the momentum range used in this analysis.  
 The uncertainty on the single-track reconstruction efficiency is 5\% and is correlated point-to-point where it contributes to the scale uncertainty in the correlation functions and yields.

The BEMC is used for the neutral-energy reconstruction and triggering.  It is a lead-scintillator sampling calorimeter with full $2\pi$ azimuthal coverage and a pseudorapidity range of $|\eta| < 1.0$.  The BEMC has 4800 towers with a transverse size of $0.05 \times 0.05$ in azimuth $\phi$ and pseudorapidity $\eta$.  This analysis uses events triggered by a high-energy deposit in a BEMC tower, referred to as a `High Tower' (HT).  The raw trigger threshold corresponds to  approximately 5.4 GeV of transverse energy ($\Et$).  Only towers above $\Et >$ 2.0 GeV are used in this analysis for jet reconstruction.  This energy threshold excludes minimally ionizing particles.  Partially formed hadronic showers may still pass this threshold and deposit charged energy.  Double counting of charged hadrons is avoided by correcting the tower energies as in Refs.~\cite{Abelev:2013fn,PhysRevLett.119.062301}.  This is especially important during the jet-finding procedure when neutral constituents are included in jet reconstruction~\cite{PhysRevLett.115.092002}.  When a tower has tracks matched to it, the tower energy is adjusted by:

\begin{equation}
\Delta E_{\rm corr} = \begin{cases}
    E_{\rm tow} & \text{for  } E_{\rm tow} < f\times \sum\limits_{\rm matches} p \\
    \\
    f\times \sum \limits_{\rm matches} p & \text{for  } E_{\rm tow} > f\times \sum \limits_{\rm matches} p .
\end{cases}
\label{hadcorr_eqn}
\end{equation}

\noindent where $E_{\rm tow}$ is the tower energy and $\sum\limits_{\rm matches} p$ corresponds to the total momentum magnitude summed over all matching tracks. The fraction $f$ is chosen to be 1 in order to remove 100\% of the deposited charged energy. The tower is corrected by assigning new energy $E_{\rm new} = E_{\rm tow} - \Delta E_{\rm corr}$ to the tower. However, the tower is discarded when the new energy is below the 2.0 GeV threshold required for jet reconstruction. 

\section{Analysis Method} \label{sect:Method}
This measurement utilizes data collected during the 2014 run from \AuAu collisions at nucleon-nucleon center-of-mass energy of \sNN = 200 GeV by the STAR experiment~\cite{Ackermann:2002ad} at RHIC.  Events referred to as signal events are required to contain a HT trigger in the BEMC \cite{Beddo:2002zx}.  Minimum-bias (MB) triggered events based on coincidence of Zero Degree Calorimeters (ZDC coincidence), Vertex Position Detectors and Beam-Beam Counters signals are used to estimate the pair-acceptance effects via a mixed-event (ME) technique \cite{PhysRevC.96.024905}. For this analysis, 9.4M HT-triggered and 4.0M MB collision events are used.  Events are further categorized by their centrality selection, defined in section ~\ref{sect:CentEP}.  The events are required to have a reconstructed primary vertex $|v_{z}| <$ 24 cm and centrality of 20-50\%.  

The reaction plane is approximated by the second-order event plane, which is the experimentally reconstructed second-order symmetry plane, and will be referred to as the ``event plane" ($\Psi_{2,EP}$) in this text, for simplicity.

The distributions of these associated tracks relative to the trigger jet are measured in three bins in the angle between the trigger jet and the event plane, in-plane ($| \Psi_{2,EP}-\phi_{\mathrm {jet}}|<\pi/6$), mid-plane ($\pi/6<|\Psi_{2,EP}-\phi_{\mathrm {jet}}|<\pi/3$), and out-of-plane ($|\Psi_{2,EP}-\phi_{\mathrm {jet}}|>\pi/3$) bins.  The analysis is restricted to 20--50\% central \AuAu collisions to achieve the highest event-plane resolution and therefore the analysis will be most sensitive to any path-length dependencies. 

\subsection{Centrality determination and event plane reconstruction} \label{sect:CentEP}

Centrality is a measure of the transverse overlap between the colliding nuclei and is generally expressed as a percentage of all collisions. 
For example, the 0-10\% most central events would refer to the 10\% of events with the most overlap and thus the 10\% smallest impact parameter.  This analysis studied semi-peripheral (20-50\%) events to maximize the eccentricity of the interaction region.  Centrality is determined by fitting the  charged-particle multiplicity from the TPC within $|\eta| <$ 0.5 that is corrected for dependence on the $v_z$ and the beam luminosity.

\label{sect:RxnPlane}
Within the overlap region, symmetry planes are generated from initial asymmetries in the nucleon distributions and can be quantified by a harmonic decomposition ~\cite{Poskanzer:1998yz}.  The reaction plane would correspond to the second-order symmetry plane $\Psi_{2, EP}$ if nucleon distributions were in their average positions and devoid of fluctuations of interactions amongst nucleons ~\cite{PhysRevC.101.064901}.  We refer in this letter to the event plane as being the experimentally reconstructed second-order symmetry plane ~\cite{Poskanzer:1998yz}.

By measuring the charged particle azimuthal distribution, the n-th order event plane can be extracted by \cite{Poskanzer:1998yz}:  

\begin{equation}
 \Psi_{n,EP} = \frac{1}{n} \arctan \left(\frac{Q_{y, n}}{Q_{x, n}} \right),
 \label{eqn:EP}
\end{equation}

Where, the weighted $Q$ vectors are given by:

\begin{align}
    Q_{x, n} = \sum_{\mathrm{tracks}} w_{\rm track} \cos (n \phi_{\rm track}) \notag\\
    Q_{y, n} = \sum_{\mathrm{tracks}} w_{\rm track} \sin (n \phi_{\rm track}).
    \label{Eqn:Qvecs}
\end{align}

\noindent where the sum is calculated for all
reconstructed charged particles (tracks) in the event, $\phi_{\rm track}$ is the track's azimuthal angle, and $w_{\rm track}$ the weight associated with the track.  Weights, $w_{\rm track}$, are optimized to calculate the event-plane vector to the best accuracy. This work uses the common approach of scaling by the track's $p_{\mathrm{T}}$ ($w_{\rm track} = p_{\rm T, track}$)~\cite{Poskanzer:1998yz}.  The event plane is calculated event-by-event, following a procedure similar to Ref.~\cite{CASTILHO2018369} using charged tracks with $0.2 < p_{\rm  T, track} < 1.0$ \GeV measured within the TPC.  The approach is called the Modified Reaction Plane (MRP) method \cite{Agakishiev:2014ada}.  Additional details can be found in Refs.~\cite{CASTILHO2018369,Adams_2005}.

The impact of highly energetic jets on the calculation of the event-plane orientation is reduced by removing the particles within the pseudorapidity strip ($|\deta|<0.4$) across \dphi surrounding the leading jet.  This procedure also removes a significant portion of the away-side jet, located opposite in azimuth. 
An upper limit of 1.0 GeV/c is used in the calculation of the event plane to exclude the momentum range of particles used in correlation functions from the calculation of the event plane which is used to characterize the near-side jets. Due to finite acceptance and multiplicities, the calculated event plane has an underlying anisotropy that is corrected by applying two separate correction methods.  

First, a calibration and recentering correction procedures are applied to remove bias introduced by non-uniform acceptance of the TPC tracking system and further account for potential beam-condition effects\cite{Barrette:1997pt,PhysRevC.55.1420,Poskanzer:1998yz}.  This procedure involves recentering the weighted $Q$-vectors such that, $\langle Q_{x, n} \rangle = 0 = \langle Q_{y, n} \rangle$.  

Recentering is done by calculating a modified $Q$-vector, obtained by subtracting an event averaged $Q$-vector from each event's nominal $Q$-vector and done for 10\% centrality intervals and 4 cm $v_z$ intervals.  The recentering approach, which drastically improves the uniformity of the event plane, is however, unable to remove the higher harmonics of $\Psi_{n,EP}$ \cite{Poskanzer:1998yz}.  To help remove higher harmonics and make the event-plane angle isotropic in the lab frame \cite{Voloshin:2008dg}, a second correction step, referred to as shifting, is applied event-by-event.  This method defines a new angle:

\begin{align}
\Psi'_{2,EP} =& \Psi_{2,EP} + \sum_n \frac{2}{n}
(- \langle \sin (n\Psi_{2,EP}) \rangle \cos(n\Psi_{2,EP})\notag\\& + \langle \cos(n\Psi_{2,EP})\rangle \sin(n\Psi_{2,EP})) ,
\label{eqn:Shift}
\end{align}

\noindent where the brackets denote an average over events.  We require the vanishing of each n-th Fourier moment up to 20th order.  Similarly to recentering, the shifting correction is done separately for 10\% centrality intervals and 4-cm $v_z$ intervals.  Additional details of the recentering and shifting corrections can be found in Refs.~\cite{Wang:2012bga,Barrette:1997pt,Poskanzer:1998yz}

The resulting azimuthal anisotropy can be characterized by the Fourier decomposition of the azimuthal particle distribution with respect to the second-order event plane ~\cite{Bielcikova:2003ku,Voloshin:1994mz}:

\begin{equation}
  \frac{dN}{d(\phi - \Psi_{RP})} = \frac{N_0}{2\pi} \left( 1 + 2 \sum_{n=1}^{\infty} v_n \cos[n(\phi - \Psi_{RP})] \right),
  \label{fourier_azi_part_distr}
\end{equation}

\noindent where $N_0$ is the number of particles, $\phi$ describes the azimuthal angle of the particles, $\Psi_{RP}$ describes the azimuthal angle of the true reaction plane determined by the beam axis and the impact parameter and $v_n$ is the n-th harmonic (flow) coefficient. $\Psi_{RP}$ is not experimentally known and is replaced by the reconstructed event-plane angle.  Due to finite event multiplicity, there will be a difference between these two planes.  It is quantified by event-plane resolution, $R_n$ or $R_n\{\Psi_{2, EP}\}$ given by Eqn.~\ref{resolution}. The observed $v_n$, $v_n^{\rm obs}$ is corrected for this limited resolution by doing, $v_n = v_{n}^{\rm obs}/R_n$ \cite{Abbas:2013taa,Poskanzer:1998yz}.  Because an ideal event-plane resolution is equal to 1, for non-ideal cases, the value of the coefficients will be raised by applying the correction. Thus, $R_n$ impacts the flow-modulated background for these correlations, as described in Sect.~\ref{sect:CorrBkgd},

\begin{equation}
  R_n = R_n\{\Psi_{2, EP}\} = \langle \cos(n[\Psi_{RP} - \Psi_{2, EP}]) \rangle \quad \text{$n$ is even}
  \label{resolution}
\end{equation}

Furthermore, individual events are divided into two random sub-events by assigning charged tracks to sub-events, ``$a$" and ``$b$". The sub-events are unique, with approximately equal multiplicities. We can write the correlation of two event planes by taking the product of two sub-events \cite{Barrette:1996rs,Poskanzer:1998yz}:

\begin{align}
  \langle \cos(n[\Psi_{2, EP}^{a} - \Psi_{2, EP}^{b}]) \rangle & = \langle \cos(n[\Psi_{2, EP}^{a} - \Psi_{RP}]) \rangle\notag\\ 
         &\langle \cos(n[\Psi_{2, EP}^{b} - \Psi_{RP}]) \rangle,
  \label{resolution2sub}
\end{align}

\noindent This allows calculation of the event-plane resolution directly from data. Since $a$ and $b$ have equal multiplicities, the total event-plane resolution can be calculated from the correlation between the two sub-events as \cite{Poskanzer:1998yz}:

\begin{equation}
  R_n\{\Psi_{2, EP}\} = 
         \sqrt{2 \langle \cos(n[\Psi_{2, EP}^{a} - \Psi_{2, EP}^{b}]) \rangle} .
         \label{resolutionIndiv}
\end{equation}

The event-plane resolution is multiplicity dependent and calculated for separate ranges of collision centrality using the two sub-events method.  Narrower bins are calculated and then combined accordingly to match the ranges used by this analysis, by averaging the results from the narrow bins weighted by the multiplicity of each bin \cite{Voloshin:2008dg}.

\begin{figure}[!htbp]
  \begin{center}     

    \includegraphics[width=\linewidth]{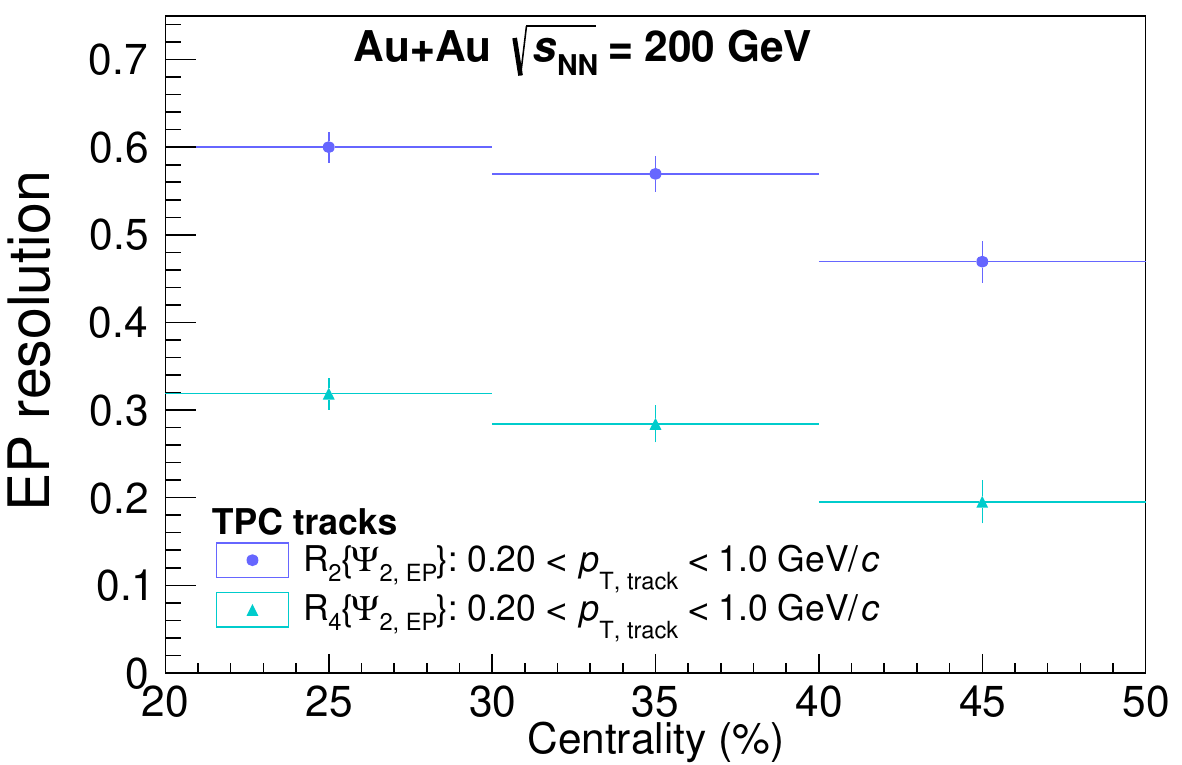}
    \caption{Event-plane resolution: Second-order (fourth-order) harmonic relative to the event plane, $R_{2}(\Psi_{2})$ ($R_{4}(\Psi_{2})$), respectively.  The approach follows the Modified Reaction Plane (MRP) method \cite{Agakishiev:2014ada} utilizing the charged tracks of the TPC for event-plane reconstruction and resolution calculation for tracks ranging from 0.2-1.0 \GeV. }  
    \label{fig:EPres}
  \end{center}
\end{figure}

The second- (fourth-) order event-plane resolutions relative to the second-order event plane ($R_2\{\Psi_{2, EP}\}$ and $R_4\{\Psi_{2, EP}\}$ respectively) as a function of collision centrality are shown in Fig.~\ref{fig:EPres}.  The resolution is peaked around the 20-30\% and 30-40\% centrality.  

The event-plane resolutions $R_2\{\Psi_{2, EP}\}$ and $R_4\{\Psi_{2, EP}\}$ were 0.56 and 0.28, respectively for the 30-40\% centrality events.  The errors on the event-plane resolution calculation were less than 1\%, leading to a negligible effect on the final \dphi correlations.  Measured values for the event-plane resolution are in good agreement with prior STAR studies \cite{Agakishiev:2014ada}.  These resolutions are evaluated to correct the observed flow coefficients which arise in the fits of the combinatorial background discussed in Sect.~\ref{sect:CorrBkgd}.

\subsection{Jet reconstruction and selection} \label{sect:JetReco}
Full jets are reconstructed by measuring charged-tracks in the TPC and collecting neutral-particle information from the BEMC.  The anti-\kt algorithm \cite{Cacciari:2008gp} implemented through the FastJet package \cite{Cacciari:2011ma} clusters these particles into jets by reconstructing the jet momenta as the quadratic sum of their constituent momenta using a boost-invariant $\it{p}^{2}_{\rm T}$ recombination scheme.  Tracks used for reconstructing jets are assumed to be pions while the towers to have arisen from massless particles.  The location of a jet, described by the `jet axis', refers to the azimuthal and pseudorapidity coordinates of the centroid of the jet. Jets can further be described by a resolution parameter, $r$, which is an input into the anti-\kt algorithm. The $r$ parameter determines the radial extent of jet constituents about the jet axis given by $\Delta r = \max(\sqrt{\dphi^2 + \deta^2})$ where \dphi (\deta) is the azimuthal angle (pseudorapidity) of constituents relative to the jet-axis. All jets measured in this work are clustered with resolution parameter of $r=0.4$. The area of jets, $A_{\rm jet}$, is found with FastJet using active ghost particles \cite{Cacciari:2008gn}.  Partially reconstructed jets at the edge of the acceptance are rejected by applying a fiducial cut, $|\eta_{\rm jet}| < 1.0 - r$, to assure all jets fall within the acceptance of the detectors.

Jets produced in heavy-ion collisions sit on top of a large amount of underlying event.  The jet signal can be found beneath tens to hundreds of particles resulting from various other processes. To reduce the influence of these background particles, this analysis requires tracks (towers) with $\pt (\Et) >$ 2.0 \GeV for jet reconstruction.  At RHIC energies this selection reduces the median background energy density per unit $A_{\rm jet}$, $\langle \rho \rangle$, down to $\approx 0.6$ GeV.  This high-constituent selection, referred to as a ``hard-core" jet selection reduces fluctuations, fake jets, and background jet particles~\cite{PhysRevLett.119.062301}.  To further reduce contributions from the background and to match the trigger condition, the jets are required to contain a constituent tower that fired the HT trigger ($ E_{\rm T, tower}> 5.4$ GeV) and a track with $p_{\rm T, track} > 4$ \GeV. The remaining jets are studied in classes of jets with \pttrigrange{15}{20} and \pttrigrange{20}{40}.

\subsection{Jet-hadron correlation} \label{sect:Correlations}
The measurement of the correlation function (distribution of charged hadrons relative to reconstructed jets) described in Eqn.~\ref{Eq:JHCorrIntro} requires several corrections.  The correlation function is measured in pseudorapidity (\deta) and azimuth (\dphi) as:\ 

\onecolumngrid\
\begin{align}
 \frac{1}{N_{\rm trig}} \frac{d^2N_{\rm assoc,jet}}{d\dphi d\deta} = \frac{1}{N_{\rm trig}} \frac{1}{\epsilon(\ptAssoc, \eta_{\rm assoc})} \frac{1}{a(\ptAssoc, \text{\dphi}, \text{\deta})}\left(\frac{d^2 N^{\rm meas}_{\rm assoc,jet}}{d\dphi d\deta} - \frac{d^2 N^{\rm bkgd}_{\rm assoc,jet}}{d\dphi d\deta}\right) .
 \label{Eqtn:CorrFunction}
\end{align}
\twocolumngrid\

 $N_{\rm assoc, jet}$ gives the number of pairs of trigger jets and the associated hadrons, $N^{\rm meas}_{\rm assoc,jet}$ and $N^{\rm bkgd}_{\rm assoc,jet}$ are the number of pairs measured and the pair characterized as background respectively.\ $\epsilon(\ptAssoc, \eta_{\rm assoc})$ is the single-track reconstruction efficiency
 .  The pair acceptance, $a(\ptAssoc, \text{\dphi}, \text{\deta})$, is calculated from the raw pairs that we measure from a trigger jet associated with charged hadrons from mixed events.  

The correlations are determined in bins of centrality, reconstructed trigger-jet transverse momentum (\ptJet), associated-hadron transverse momentum (\ptAssoc), and bins of the trigger jet relative to the event plane (in, mid, out, all combined angles) defined in Sect.~\ref{sect:Method}.  The corrected correlation functions contain a large combinatorial background ($N^{\rm bkgd}_{\rm assoc,jet}$), which must be subtracted.  This subtraction procedure is described in Sect.~\ref{sect:CorrBkgd} .

\label{sect:MixedEvents}

The pair-acceptance correction, $1 / a(\ptAssoc, \text{\dphi}, \text{\deta})$, accounts for the finite acceptance of the TPC and kinematic selections imposed on jets and tracks used in correlations, and is found by correlating jets from HT-triggered events with associated hadrons from MB events of the same event class. 
In addition to providing the acceptance correction, the mixed-event procedure will also help remove the trivial correlation due to an $\eta$ dependence in the single-particle track distributions~\cite{PhysRevC.101.064901}.  The pair acceptance will serve as the dominant effect, given that there is little $\eta$ dependence in both the tracks and jets across the acceptance range of this analysis.

The event-mixing procedure used in this analysis is well described in Ref.~\cite{PhysRevC.96.024905}.  The mixed events used for calculating $a(\ptAssoc, \text{\dphi}, \text{\deta})$ in this work are required to be within the same 10\% centrality class and to have a vertex position within 4 cm along the direction of the beam ($v_{z}$).  They are constructed separately for 20-30\%, 30-40\%, and 40-50\% centrality classes and combined accordingly.  High-momentum tracks are nearly straight, so the detector acceptance does not change significantly at high-\ptAssoc, and thus all associated momentum bins greater than 2.0 \GeV are combined to increase statistics.  There is no difference in efficiency and acceptance within uncertainties for different orientations of the jet relative to the event plane, and therefore the same correction $a(\ptAssoc, \text{\dphi}, \text{\deta})$ is applied for all angles relative to the event plane.  The acceptance $a(\ptAssoc, \text{\dphi}, \text{\deta})$ is normalized to 1 at its maximum, determined using the region of approximately constant acceptance ($|\deta| <$ 0.4).  For each \ptAssoc bin, the projection of the flat plateau region (integrated over the region $|\deta| <$ 0.4) was fit with a constant over the whole \dphi range.  The associated uncertainties of the fits were used for the systematic uncertainty on the mixed-event normalization, which is added in quadrature and reported as the scale uncertainty.  This systematic uncertainty is under 1\% (1.25\%) in all \ptAssoc bins used for the reported results of jets with  \pttrigrange{15}{20} and  \pttrigrange{20}{40}.

\subsection{Flow modulation of combinatorial background} \label{sect:CorrBkgd}
The combinatorial background ($N^{\rm bkgd}_{\rm assoc,jet}$) from Eq.~\ref{Eqtn:CorrFunction} are parametrized for trigger jets restricted to the orientation $\Re = {\text{in/mid/out-of-plane}}$ relative to the event plane in Eq.~\ref{Eqtn:JBBBCorrelations}~\cite{Bielcikova:2003ku,Sharma:2015qra}, where $v_{n,\mathrm{assoc}}$ and $v^\Re_{n, \mathrm{jet}}$ are the Fourier coefficients of the azimuthal angle distribution (flow coefficients) of background associated particles and trigger jets restricted to the orientation $\Re$ respectively. $B^\Re$ gives the background level amplitude for orientation $\Re$. The trigger jets being restricted to orientation $\Re$ modifies their nominal flow coefficients $v_{n, \mathrm{jet}}$'s into the $v^\Re_{n, \mathrm{jet}}$ according to Eq.~\ref{Eq:vRn}~\cite{Bielcikova:2003ku}. \noindent Where, $\beta^\Re$ is the $\Re$ dependence of $B^\Re$ given by Eq.~\ref{Eqtn:B_R}, where $\phi_S^\Re$ and $c$ and are the center and width of the $|\Psi_{2,EP}-\phi_{\mathrm {jet}}|$ range for jets restricted to orientation $\Re$.\

\onecolumngrid\
\begin{align}
		\left(\frac{1}{\pi}\frac{dN^{\rm bkgd}_{\rm assoc,jet}}{d\dphi}\right)_\Re = B^\Re\left( 1 + \sum_{n=2}^{\infty} 2 v_{n,\mathrm{assoc}} v^\Re_{n, \mathrm{jet}} \cos(n\dphi)\right)
		\label{Eqtn:JBBBCorrelations},
\end{align}
\begin{equation}
  v^{\Re}_{n, \mathrm{jet}} = 
  \begin{cases}
  \frac{1}{\beta^\Re}\left(v_{n, \mathrm{jet}}+\cos(n\phi_{s}^\Re)
  \frac{\sin(nc)}{nc}R_{n} + \sum_{k=2,4,6,..} (v_{(k+n), \mathrm{jet}}+v_{|k-n|, \mathrm{jet}})
  \cos(k\phi_{s}^\Re) \frac{\sin(kc)}{kc}R_{n}\right)& \text{if $n$ is even,}\\
  v_{n, \mathrm{jet}} & \text{if $n$ is odd.}
  \end{cases}
  \label{Eq:vRn}
\end{equation}
\begin{equation}
    B^\Re \propto \beta^\Re = 1 + \sum_{k=2,4,6,..} 2v_{k, \mathrm{jet}} \cos(k\phi_{s}^\Re) \frac{\sin(kc)}{kc}R_{n}
    \label{Eqtn:B_R}
\end{equation}
\twocolumngrid\

Odd $v_{n, \mathrm{jet}}$'s mainly arise from initial state fluctuations and are therefore uncorrelated with the second-order event-plane and remain constant when the trigger jet is moved relative to the event plane, while even $v_{n, \mathrm{jet}}$'s will change~\cite{Aad:2014fla,Sharma:2015qra}. The $\Re$ dependent background shape is dependent upon the event-plane resolution ($R_{n}$), which is fixed at the measured values. Extended details into the derivations of relevant equations can be found in Refs.~\cite{Bielcikova:2003ku,Nattrass_2018,Nattrass_Reexam2018}. 

Collective particle flow plays a major role in understanding the underlying-event background.  This background consists of particles created from mechanisms unrelated to the hard process that led to a jet.  Some of the jet signal is correlated with our event plane due to the path-length dependence of partonic energy loss, while soft hadrons are predominantly correlated with the event plane due to hydrodynamical flow that also contribute to the bulk-particle production. 

To remove the combinatorial background comprised by contributions from the underlying event, the reaction plane fit (RPF) developed in Ref.~\cite{Sharma:2015qra} is applied in this work. The measured jet-hadron correlation signal is decomposed into a \ns and an \as, with the former being narrow in \dphi and \deta and the latter narrow in \dphi, but broad in \deta.  The narrowness of the near-side implies the signal is negligible at large \deta, where the background dominates.  Applying RPF, we defined our `signal + background' region for $|\deta| < 0.6$ and further $0.6 \leq |\deta| < 1.2$ as a background dominated region where the signal was assumed to be negligible.  The correlation function is fit with the parametrization given in Eq.~\ref{Eqtn:JBBBCorrelations}, restricted to $n=4$ for the background dominated region at large \deta and small \dphi ($|\Delta\phi|<\pi/2$) simultaneously for in-plane, mid-plane, and out-of-plane trigger jets. RPF improves upon prior background subtraction techniques~\cite{PhysRevC.101.064901,Nattrass:2016cln} by avoiding problems due to contamination from jets on both the near- and away-side by using the \ns at large \deta and also using the dependence of the flow-modulated background on the angle of the trigger jet relative to the event plane to constrain the background shape and level.

For $\ptAssoc >$2 \GeV, the combinatorial background is small, and few high momentum tracks are found at large distances in pseudorapidity from the near-side jet. So the model fit is restricted to $n = 3$ as higher order terms are no longer contributing.  This occurs for $\ptAssoc >$ 4 \GeV with \pttrigrange{15}{20} jets and $\ptAssoc >$ 3 \GeV with \pttrigrange{20}{40}.  Therefore, the RPF fits consist of six parameters ($B^\Re$, $v^\Re_{2, \rm jet}$, $v_{2, \mathrm {assoc}}$, $(v_{3, \mathrm {jet}}\times v_{3, \mathrm {assoc}})$, $v^\Re_{4, \mathrm {jet}}$, and $v_{4, \rm assoc}$) for low \ptAssoc, and four ($B^\Re$, $v^\Re_{2, \rm jet}$, $v_{2, \mathrm {assoc}}$ and $(v_{3, \mathrm {jet}}\times v_{3, \mathrm {assoc}})$) for high \ptAssoc.\ 

\begin{figure}[!htb]
\centering
 \includegraphics[width=0.9\linewidth]{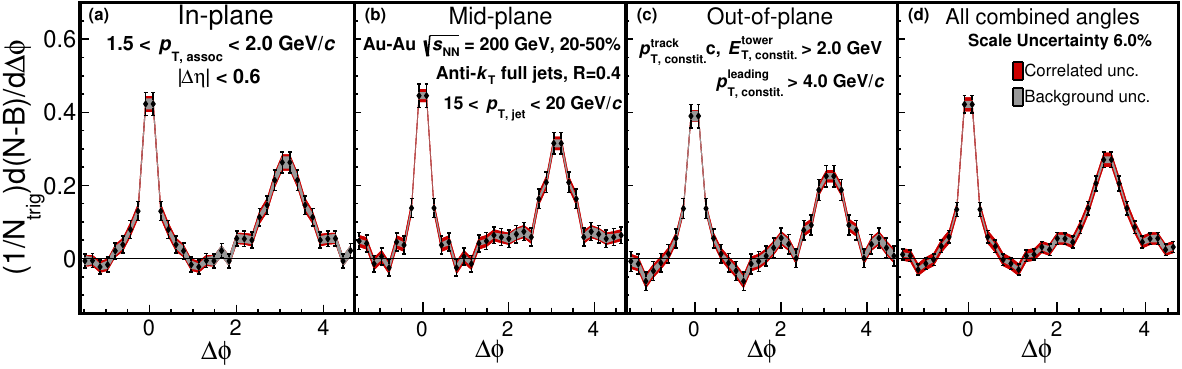}
\caption{Example of a background subtracted correlation function for \pttrigrange{15}{20} jets correlated with \ptassocrange{1.5}{2.0} charged hadrons from the 20-50\% most central collisions.   Correlated scaled uncertainty from the application of mixed events is displayed by the red band, while the uncertainty associated with the RPF background fit is displayed as the gray band. }
\label{plot:dPhi1.5-2.0}
\end{figure}

Comparison of in-, mid-, and out-of-plane jet-hadron correlations is performed after background subtraction to explore the effects related to event-plane orientation.\ An example of event-plane dependent correlation function after the RPF background subtraction for jets with \pttrigrange{15}{20} is shown in Fig.~\ref{plot:dPhi1.5-2.0} for in-plane (a), mid-plane (b), out-of-plane (c) and jets from all combined angles (d) for associated particles with momenta \ptassocrange{1.5}{2.0}.  The uncertainties from the RPF background subtraction are propagated using the covariance matrix from the fit and are non-trivially correlated point-to-point and between different bins relative to the event plane.  These are shown in gray.  The uncertainty from the acceptance correction, described in Sect.~\ref{sect:MixedEvents} is displayed by the red uncertainty band. The uncertainties on the event-plane resolution are negligible relative to that of the background subtraction and statistical uncertainties of the final results.  Additional uncertainties uncorrelated with each other, but correlated for all points in \dphi are given in Tab.~\ref{Tab:SystematicSummary}.  They are combined in quadrature and listed as the scale uncertainty on the results.\

\onecolumngrid\
\begin{figure}[!htb]
  \centering   
  \includegraphics[width=0.8\linewidth]{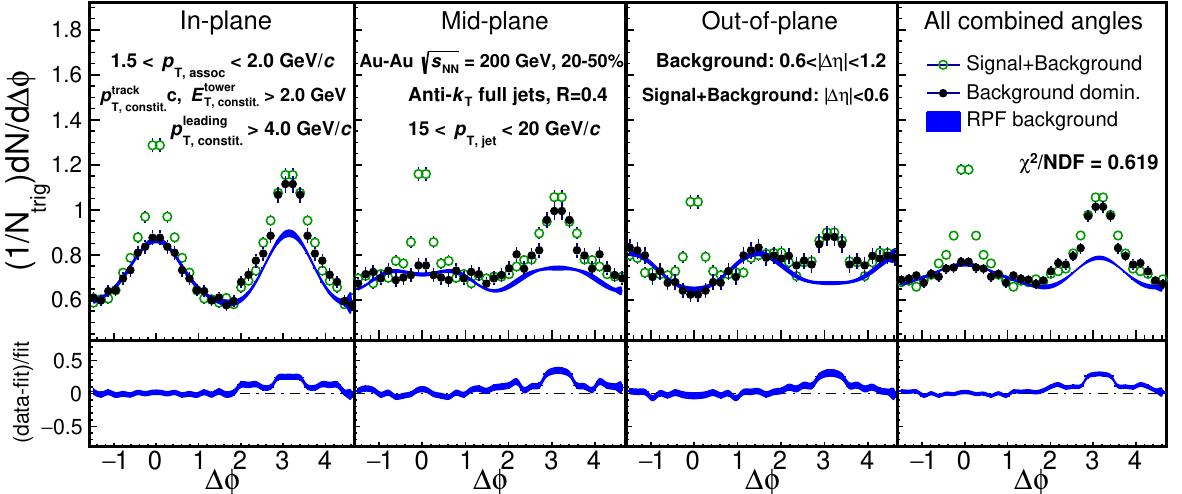}  
  \caption{(Top) Signal+background region, background-dominated region, and RPF fit to the background for the event-plane dependent jet-hadron correlations of $15 < p_{\rm T}^{\rm jet} < 20$ \GeV jets correlated with \ptassocrange{1.5}{2.0} charged hadrons from the 20-50\% most central events. (Bottom) Quality of the RPF fit to the background-dominated region expressed as (data - fit) / fit. }
  \label{plot:SigBgFit1.5-2.0ex}
\end{figure}
\twocolumngrid\

\onecolumngrid\
\twocolumngrid\

\subsection{Systematic Uncertainties} \label{sect:SysUnc}
The PYTHIA6 Perugia 2012 tune is used to create particle level dijet events embedded in MB \AuAu 200 GeV events at the detector level.  This allows for further comparisons between the jets from the input PYTHIA6 tracks (generator-level jets or GEN-jets) and the jets from the embedded tracks reconstructed by GEANT (reconstructed jets or RECO-jets). RECO-level events are analyzed with the same selections and parameters as used by the data analysis. All particles of GEN-level events are required to be in their final state (particles with no further daughters). GEN-level jets have $\ptGenJet > $ 10 \GeV and the RECO-level jets have $\ptRecoJet >$ 5 \GeV. Further, we require only the RECO-level jets contain a neutral component with $ E_{\rm T} \geq 5.4$ GeV to match the trigger condition applied in data. Our goal is to study the effects of the detector response on jet reconstruction and our analysis.  In order to compare the same jet pre- and post-reconstruction, we apply a `nearness' criteria in the $\eta-\phi$ that matches a GEN-jet to one RECO-jet satisfying: 

(a) $r_{\rm GEN, RECO}\leq0.4$, $r_{\rm GEN, RECO}$ being the separation between the gen-jet and the reco-jet in $\eta-\phi$ space, (b) the RECO-jet is the closest to the gen-jet among all reco-jets (minimize $r_{\rm GEN, RECO}$).  

We compare the resulting GEN- and RECO-level jet spectra to calculate the momentum resolution which is given by:

\begin{equation}
    \ptJet \textnormal{ resolution} (\%) = \frac{\text{\ptGenJet} - \text{\ptRecoJet}}{\text{\ptGenJet}} \times \text{100} \%.
\end{equation}

\begin{figure}[!htbp]
	\centering
    \includegraphics[width=0.8\linewidth]{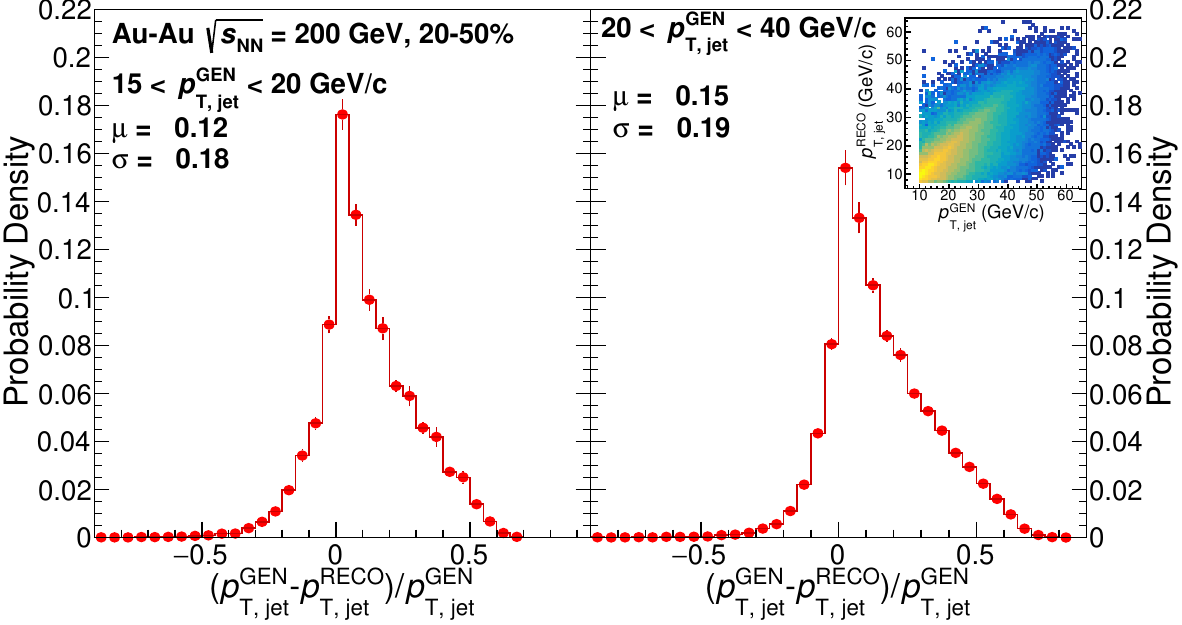}
	\caption{\ptJet resolution for $15 < \ptGenJet < 20$ \GeV (left) and $20 < \ptGenJet < 40$ \GeV (right) $R=0.4$ full jets with the corresponding response matrix as an inset on the top right.  Jets are measured from all angles relative to the event plane in the 20-50\% most central events.  This comparison is for matched GEN-level to RECO-level jets. }
	\label{fig:jetptresWinset}
\end{figure}

Fig.~\ref{fig:jetptresWinset} shows the \ptJet resolution for $15 < \ptGenJet < 20$ \GeV (left) and $20 < \ptGenJet < 40$ \GeV (right) $R=0.4$ full jets.  The distributions have been normalized into probability functions and are shown for the 20-50\% most central events for all angles of the jet relative to the event plane.  Fig.~\ref{fig:jetptresWinset} further shows an average energy loss of around 15\% going from GEN to RECO. This net energy shift is thought to be due to counteracting effects of the tracking inefficiency at the RECO level and there being more tracks in the RECO level from the min-bias pedestal. The \ptJet resolution for $15 < \ptRecoJet < 40$ \GeV reco-jets is roughly 10-20\%.  The event-plane dependence of the \ptJet resolution was also studied and found to be within 1-2\% of each other between different orientations of jets with respect to the event plane.  There can be slight differences in the jets reconstructed at lower momenta with \pttrigrange{15}{20} for jets at different angles relative to the event plane, due to a low momentum embedded jet overlapping with another jet in the \AuAu data and from statistical fluctuations.  As there are more jets from in-plane than out-of-plane orientations in the data, this leads to an apparent difference in the reconstructed jet spectra. Otherwise there are no significant differences between jets at different angles relative to the event plane. A map between \ptGenJet and  \ptRecoJet, called the response matrix, is added to Fig.~\ref{fig:jetptresWinset} as a smaller inset on the top right.\   
\onecolumngrid\
\twocolumngrid\
We summarize the systematic uncertainties in Tabs.~\ref{Tab:SystematicSummary}, \ref{Tab:SystematicSummaryEPdep}, and \ref{Tab:SystematicSummaryEPdep2040}.  Table~\ref{Tab:SystematicSummary} lists the sources of systematic uncertainties which are independent of the angle relative to the event plane.\  These sources include the single-track reconstruction efficiency (Sect.~\ref{sect:STARdet})
 and uncertainties in the event-plane resolution (Sect.~\ref{sect:CentEP}).  

There is a shape uncertainty associated with the application of the acceptance correction due to slight changes in the correlation function at large \deta in the acceptance with $v_z$ position.  The background level is determined from the level of the correlation function at large \deta, leading to a scale uncertainty in the background subtraction dependent on \ptAssoc.  This uncertainty is from the differences between the nominal (unbinned in $v_{z}$ or $v_{z}$-integrated) and the $v_{z}$-binned method for correcting the mixed events on the level of the background in the $0.6 < \deta < 1.2$ range, and signal plus background, in the $|\deta| < 0.6$ range.  The large \deta region is used to determine the background, so any uncertainties in the level of the correlation function in this region lead to an uncertainty in the level of the background in the signal region.  This is expressed as an additional scale band on the final results.  This uncertainty is determined by varying the binning of the mixed events in $v_z$ and is correlated for different angles relative to the event plane and for different bins in \ptAssoc.  Above $\ptAssoc >$ 3 \GeV, this uncertainty is negligible because the background is small. 

A shape uncertainty in \dphi due to the $v_{z}$-binning, similar to that in \deta, could lead to an additional uncertainty in the correlation functions.  To test for such uncertainty, the ratio of the 1D \dphi projection calculated with the nominal method and the $v_{z}$ binned method with 4cm $v_z$ bins was calculated for each \ptJet, \ptAssoc, and centrality bin. The variations are smaller than the statistical errors associated with the points. This uncertainty was therefore considered negligible.

The uncertainties are added in quadrature and lead to a 6\% uncertainty in the scale of the correlation functions and yields with the single-track reconstruction efficiency being the dominant source.  This uncertainty is uncorrelated for different associated-particle momenta.

\begin{table}[!htbp]
\caption{Summary of systematic uncertainties which are independent of the angle relative to the event plane and the momentum for both \pttrigrange{15}{20} and \pttrigrange{20}{40} in 20-50\% central \AuAu collisions.}
\label{Tab:SystematicSummary}
\centering 
\begin{tabular}{l | c} 
\hline 
Source & Uncertainty \% \Tstrut \Bstrut \\ 
\hline 
Single-particle reconstruction efficiency & 5 \\
\hline
Mixed-event (shape \dphi) & negligible \\
\hline
Mixed-event normalization & $< 1.25$ \\ 
\hline
Event-plane resolution &  $< 1$ \\ 
\hline
\end{tabular}
\end{table}

Additional uncertainties, highly dependent on the angle of the jet relative to the event plane and the associated particles' momentum, are summarized in Tabs.~\ref{Tab:SystematicSummaryEPdep} and~\ref{Tab:SystematicSummaryEPdep2040}.  These include the impact of the scale uncertainty from the mixed events (Sect.~\ref{sect:Correlations}) and the RPF background fit (Sect.~\ref{sect:CorrBkgd}) on the associated yield and jet-peak width results.  These uncertainties are compared for two associated particle momentum bins (1.0-1.5 and 3.0-4.0 \GeV) to highlight how much of an impact the background has at low momenta.  As indicated in Table 3, the uncertainties are considerably larger for $p_T$-associated tracks when $p_T < 2 \, \text{GeV}/c$. This arises from the reduced jet-associated track yields in the $1.5-2 \, \text{GeV}$ range and increased background levels, leading to a less accurate fit. This interplay becomes more pronounced at higher jet $p_T$, where the discrepancy between tracks $< 2 \, \text{GeV}$ and tracks $> 2 \, \text{GeV}$ becomes more significant. The uncertainty arising from the RPF background subtraction is non-trivially correlated point-to-point in \dphi and for different orientations of the jet relative to the event plane, but is uncorrelated between different \ptAssoc bins.  The acceptance-correction uncertainty on the shape in \deta is also correlated for different orientations of the jet relative to the event plane and uncorrelated across \ptAssoc bins.  The acceptance-shape uncertainty is the dominant source at low-momenta, while being more comparable to the background uncertainty at larger momenta.

To calculate the jet-energy shift (JES) due to underlying-event background contribution to the reconstructed jet energy, background levels were calculated by summing over $p_{\rm T, track}$'s in random cones of radius $0.4$ in $\eta-\phi$ plane, thrown in minimum-bias events of matching centrality selection. The mean background levels are found to be $0.388$ \GeV (in-plane), $0.344$  \GeV (mid-plane), and $0.308$  \GeV (out-of-plane), with a corresponding RMS of $0.4$ \GeV. Shifting the jet-momenta selection by these mean background levels are utilized to calculate the associated systematic uncertainties.

\onecolumngrid\
\begin{table}[!htbp]
\caption{Summary of systematic uncertainties on the associated yields and widths calculated from the correlation functions due to the shape uncertainty coming from the shape of the acceptance correction in \deta, the correlated background-fit uncertainty, and the uncertainty associated with the jet energy shift (JES) correction, each varying with event-plane orientation bins.  They are displayed for \pttrigrange{15}{20} in 20-50\% central \AuAu collisions for \ptassocrange{1.0}{1.5} and \ptassocrange{3.0}{4.0} bins.  The values are expressed as a percent of the nominal value.}
\label{Tab:SystematicSummaryEPdep}
\centering
\begin{tabular}{c | c | c | c | c | c | c}\hline
  \multirow{3}{*}{Source} & \multirow{3}{*}{Result} & \multirow{3}{*}{Orientation} & \multicolumn{4}{c}{Uncertainty \%} \\ \cline{4-7}
  &   &      & \multicolumn{2}{c|}{Near-side: \ptAssoc (\GeV)} & \multicolumn{2}{c}{Away-side: \ptAssoc (\GeV)} \Bstrut \\ \cline{4-7}     
  &   &      & 1.0-1.5 & 3.0-4.0 & 1.0-1.5 & 3.0-4.0 \\ \hline       
  
  \multirow{6}{*}{} & \multirow{3}{*}{Yield} & in-plane & 14 & 1.2 & 8.1 & 3.0  \\ \cline{3-7}
  & & mid-plane & 11 & 1.2 & 7.6 & 3.3 \\ \cline{3-7}
  Acceptance & & out-of-plane & 11 & 1.1 & 7.3 & 3.0 \\ \cline{2-7}
  shape & \multirow{3}{*}{Width} & in-plane & 5.2 & 0.6 & 4.3 & 2.4 \\ \cline{3-7}
  & & mid-plane    & 5.4 & 0.5 & 3.9 & 2.2 \\ \cline{3-7}
  & & out-of-plane & 4.3 & 0.4 & 4.9 & 2.1 \\ \cline{3-7}
  \hline
  
  \multirow{6}{*}{} & \multirow{3}{*}{Yield} & in-plane & 11 & 0.8 & 6.1 & 2.1 \\ \cline{3-7}
  & & mid-plane & 7.7 & 0.9 & 5.2 & 2.5 \\ \cline{3-7}
  Background & & out-of-plane & 7.4 & 0.7 & 4.7 & 2.1  \\ \cline{2-7}
  fit & \multirow{3}{*}{Width} & in-plane & 10 & 0.1 & 8.2 & 0.4 \\ \cline{3-7}
  & & mid-plane    & 10 & 0.1 & 7.5 & 0.4 \\ \cline{3-7}
  & & out-of-plane & 8.2 & 0.1 & 9.3 & 0.4 \\ \cline{3-7}
  \hline

  \multirow{6}{*}{} & \multirow{3}{*}{Yield} & in-plane & 1.8 & 3.7 & 4.9 & 4.0  \\ \cline{3-7}
  & & mid-plane & 2.8 & 2.5 & 1.4 & 4.2 \\ \cline{3-7}
  JES & & out-of-plane & 3.7 & 3.3 & 4.8 & 4.2 \\ \cline{2-7}
  correction & \multirow{3}{*}{Width} & in-plane & 0.6 & 0.4 & 4.0 & $<$ 0.1 \\ \cline{3-7}
  & & mid-plane    & 0.2 & $<$ 0.1  & 7.9 & 1.5 \\ \cline{3-7}
  & & out-of-plane & 1.6 & 0.7 & 1.9 & 1.7 \\ \cline{3-7}
  
  \hline
\end{tabular}
\end{table}

\begin{table}[!htbp]
\caption{Summary of systematic uncertainties on the associated yields and widths calculated from the correlation functions due to the shape uncertainty coming from the shape of the acceptance correction in \deta, the correlated background-fit uncertainty, and the uncertainty associated with the JES correction, each varying with event-plane orientation bins.  They are displayed for \pttrigrange{20}{40} in 20-50\% central \AuAu collisions for \ptassocrange{1.0}{1.5} and \ptassocrange{3.0}{4.0} bins.  The values are expressed as a percent of the nominal value. }
\label{Tab:SystematicSummaryEPdep2040}
\centering
\begin{tabular}{c | c | c | c | c | c | c}\hline
  \multirow{3}{*}{Source} & \multirow{3}{*}{Result} & \multirow{3}{*}{Orientation} & \multicolumn{4}{c}{Uncertainty \%} \\ \cline{4-7}
  &   &      & \multicolumn{2}{c|}{Near-side: \ptAssoc (\GeV)} & \multicolumn{2}{c}{Away-side: \ptAssoc (\GeV)} \Bstrut \\ \cline{4-7}     
  &   &      & 1.0-1.5 & 3.0-4.0 & 1.0-1.5 & 3.0-4.0 \\ \hline           
  
  \multirow{6}{*}{} & \multirow{3}{*}{Yield} & in-plane & 35 & 1.2 & 22 & 2.5  \\ \cline{3-7}
  & & mid-plane & 39 & 1.2 & 22 & 2.2 \\ \cline{3-7}
  Acceptance & & out-of-plane & 59 & 0.9 & 37 & 1.6 \\ \cline{2-7}
  shape & \multirow{3}{*}{Width} & in-plane & 30 & 0.7 & 8.4 & 1.9 \\ \cline{3-7}
  & & mid-plane    & 22 & 0.5 & 15  & 1.7 \\ \cline{3-7}
  & & out-of-plane & 19 & 0.4 & 28  & 1.2 \\ \cline{3-7}
  \hline
  
  \multirow{6}{*}{} & \multirow{3}{*}{Yield} & in-plane & 13 & 1.0 & 8.2 & 2.1 \\ \cline{3-7}
  & & mid-plane & 14 & 0.7 & 8.0 & 1.2 \\ \cline{3-7}
  Background & & out-of-plane & 22 & 0.9 & 14 & 1.5  \\ \cline{2-7}
  fit & \multirow{3}{*}{Width} & in-plane & 13 & 0.1 & 3.7 & 0.2 \\ \cline{3-7}
  & & mid-plane    & 10  & 0.1 & 6.8 & 0.2 \\ \cline{3-7}
  & & out-of-plane & 8.6 & $<$ 0.1 & 13  & 0.1 \\ \cline{3-7}
  \hline

  \multirow{6}{*}{} & \multirow{3}{*}{Yield} & in-plane & 0.2 & 0.4 & $<$ 0.1 & 2.6 \\ \cline{3-7}
  & & mid-plane & 1.1 & 0.3 & 2.1 & 2.9 \\ \cline{3-7}
  JES & & out-of-plane & 3.1 & 0.6 & 1.4 & 3.5  \\ \cline{2-7}
  correction & \multirow{3}{*}{Width} & in-plane & 1.2 & 0.5 & 3.3 & 0.3 \\ \cline{3-7}
  & & mid-plane    & 1.2 & 0.7 & 1.3 & 1.5 \\ \cline{3-7}
  & & out-of-plane & 2.2 & 0.1 & 0.2 & 1.9 \\ \cline{3-7} 
  \hline
\end{tabular}
\end{table}
\twocolumngrid\

\section{Results} \label{sect:Results}
\label{sect:Yields}
Charged-particle yields associated with jets are found by setting the \dphi integration limits in the associated-yield formula given in Eqn. \ref{Eqtn:Yield1}. For the near-side the limits are, $a=-\pi/3$ and $b=\pi/3$, while for the away-side, we have $a=2\pi/3$ and  $b=4\pi/3$.  The integration limits in \deta are the same for both the near-side and away-side, $c=-0.6$ and $d=0.6$.

\begin{figure}[!ht]
  \begin{center}     
  \includegraphics[width=15cm]{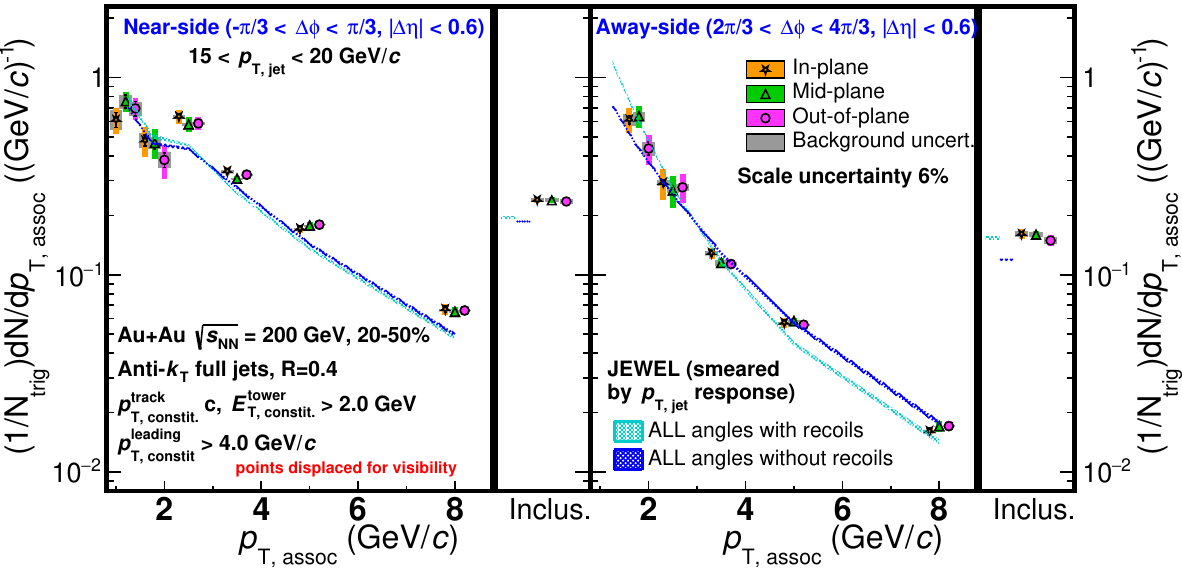}
  \includegraphics[width=15cm]{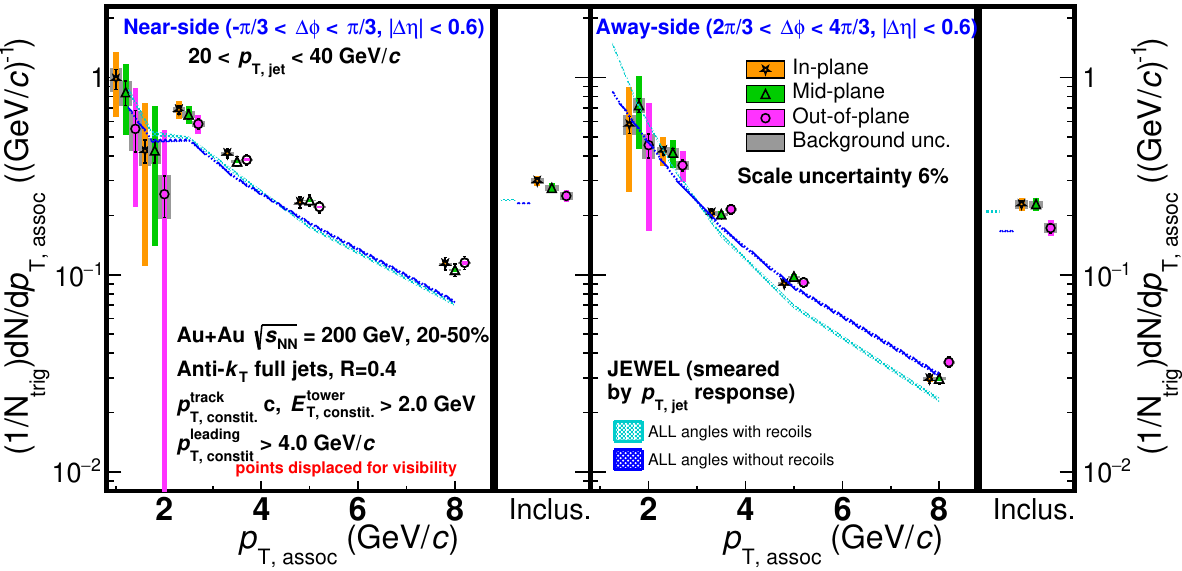}
  \caption{Near-side (left) and away-side (right) uncorrected associated yield vs \ptAssoc for 15-20 (top) and 20-40 \GeV (bottom) full jets of 20-50\% centrality in \AuAu collisions.  The grey bands describe the systematic uncertainties of the background fits and are non-trivially correlated point-to-point.  The colored bands are scale uncertainties from the mixed event acceptance shape and JES correction.  There is an additional 6\% global scale uncertainty.  Included on the far right of the near- and away-side panels is the inclusive transverse momentum bin from 1.0-10.0 \GeV.  Points are displaced for visibility.  } 
  \label{fig:AssocYields}
  \end{center}
\end{figure}

Shown in Fig.~\ref{fig:AssocYields} is the near-side (left) and away-side (right) associated yields vs \ptAssoc for \pttrigrange{15}{20} (top) and \pttrigrange{20}{40} \GeV (bottom) full jets in 20-50\% centrality collisions.  The yields are compared for each orientation of the trigger jet reconstructed relative to the event plane (in/mid/out) for \ptAssoc $\in$ [1.0, 1.5], [1.5, 2.0], [2.0, 3.0], [3.0, 4.0], [4.0, 6.0], [6.0, 10.0] \GeV. 

The main feature is the steeply falling associated yield with increasing \ptAssoc, which occurs on both the near- and away-side.  Note that associated yields for \ptAssoc $\geq$ 2.0 \GeV include jet constituents, which leads to the discontinuity at \ptAssoc = 2 \GeV.  Additionally, the use of jets with `hard-cores' and containment of a tower associated to the firing event trigger can lead to a surface bias of the near-side jet.  This, however, maximizes the average path length the away-side recoil jets travels, increasing the likelihood of an interaction with the medium. We would expect an in-plane jet and an out-of-plane jet with the same \ptJet to have different distributions of hard and soft constituents. An in-plane jet with less path travelled in the medium on average would be expected to show higher yields of constituents with higher \ptAssoc than the more quenched out-of-plane jet, which would be expected to show higher yields of constituents with lower \ptAssoc. While uncertainties are smaller for jets with \pttrigrange{15}{20}, both samples still lack any clear dependence on the event-plane angle.   This is an indication that modifications dependent on the average path length are smaller than the experimental uncertainties.  On the far right of the near- and away-side panels is the inclusive \ptAssoc bin with \ptassocrange{1.0}{10}. The associated yields of each event-plane orientation in the inclusive \ptAssoc selection are consistent with each other for the sample where \pttrigrange{15}{20}. However, in the jet sample with \pttrigrange{20}{40}, there are indications suggesting potential modifications. This potential modification is apparent on both the near- and away-side with the largest contributions coming from the lowest \ptAssoc bins.\
\onecolumngrid\
\twocolumngrid\
\label{sect:Widths}
The widths are calculated by fitting a Gaussian, $A \rm e^{(\Delta\phi - \Delta\phi_0)^2/2\sigma^2}$, to the jet peak centered at $\dphi_0 = 0$ for the near-side and $\dphi_0 = \pi$ for the away-side.\  The gaussians are fitted separately, with a range of $|\Delta\phi|~<~\pi/3$ on the \ns and $|\Delta\phi - \pi|~<~\pi/3$ on the \as. The Gaussian fit is repeated with different values of the background parameters and the covariance matrix is used to propagate the uncertainties.  The scale uncertainties on the widths are given by $\sigma_{w}^{sc} = \frac{B}{\sigma_B} \times |\alpha - 1|$, where $\alpha$ is the \pt-dependent scale factor associated with the acceptance shape when propagating the uncertainty ($\sigma_B$) associated with the background ($B$), determined in the $0.6 \leq |\deta| < 1.2$, to the region $|\deta| < 0.6$.  

\begin{figure}[!ht]
  \begin{center}     
  \includegraphics[width=16cm]{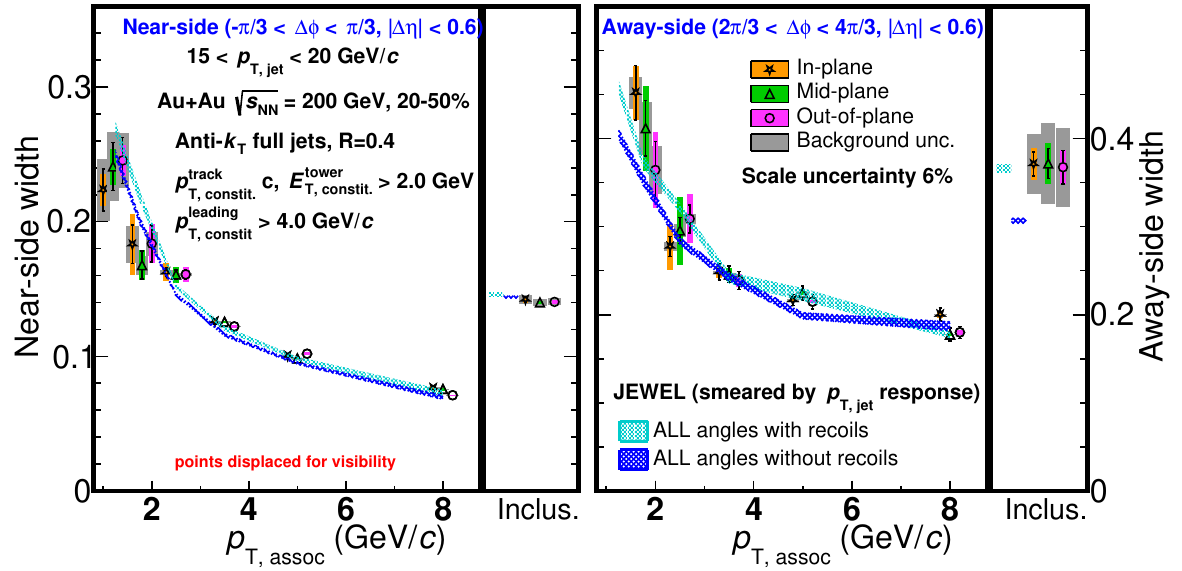}
  \includegraphics[width=16cm]{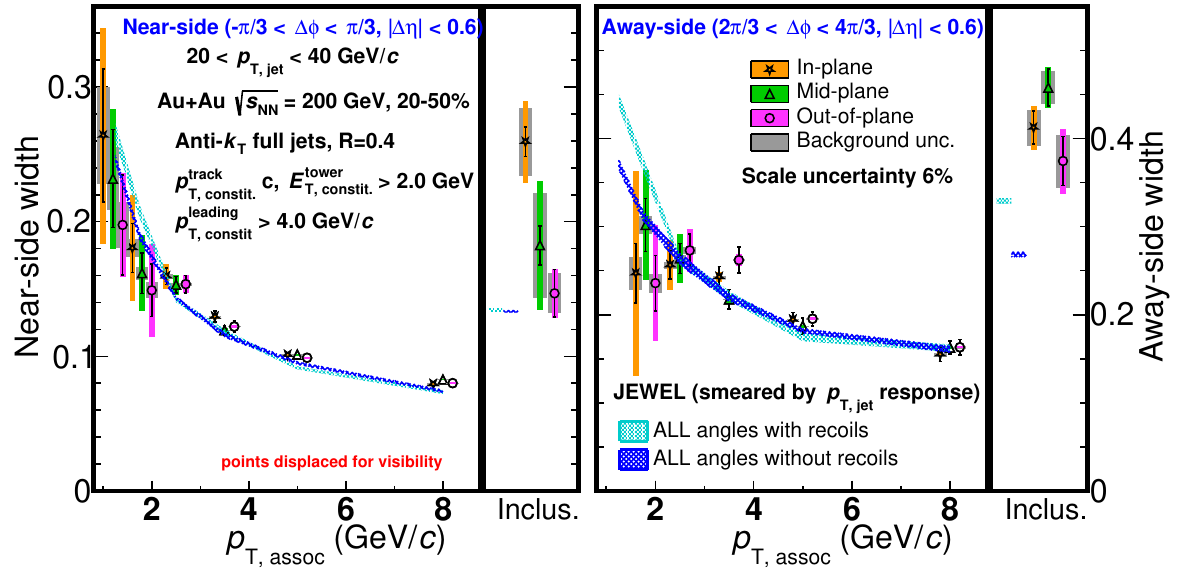}
  \caption{Near-side (left) and away-side (right) jet width vs \ptAssoc for 15-20 (top) and 20-40 \GeV (bottom) full jets of 20-50\% centrality in \AuAu collisions.  The widths are extracted from the gaussian fit to the jet peak.  The grey bands describe the systematic uncertainties of the background fits which are non-trivially correlated point-to-point.  The colored bands are scale uncertainties from the mixed-event acceptance shape and JES correction.  There is an additional 6\% global scale uncertainty.  Included on the far right of the near- and away-side panels is an inclusive transverse momentum bin from 1.0-10.0 \GeV.  Points are displaced for visibility. }
  \label{fig:JetWidths}
  \end{center}
\end{figure}

From Fig.~\ref{fig:JetWidths}, it is clear a broadening of the jet peaks is occurring for decreasing \ptAssoc.  This is expected from both collisional energy loss and gluon bremsstrahlung.  With out-of-plane jets expected to traverse a longer average path length than in-plane jets, this would lead to additional interactions with the medium and more subsequent re-scatterings resulting in a larger width for jets out-of-plane relative to in-plane.   Within the uncertainties there is no clear ordering.  This indicates the effect of path-length dependent energy loss is not large enough to be seen by the current precision of the data.  On the far right of the near- and away-side panels is the inclusive \ptAssoc bin with \ptassocrange{1.0}{10}.   The widths of each event-plane orientation in the inclusive selection for the the sample with \pttrigrange{15}{20} are consistent.  Conversely, the \pttrigrange{15}{20} sample reveals indications of potential modifications.
This potential modification is apparent on both the near- and away-side and primarily comes from the lowest transverse momentum bins where sample size is the largest.\
\onecolumngrid\
\twocolumngrid\

The measurements presented in Figs.~\ref{fig:AssocYields} and~\ref{fig:JetWidths} are compared to calculations from ``Jet Evolution With Energy Loss" known as JEWEL~\cite{Zapp_2009}, a jet energy loss model based on radiative and collisional energy loss in connection with partons sampled from a longitudinally expanding  medium~\cite{Zapp:2013vla}.\   To enhance the accuracy of the data-to-model comparison, we incorporate a smearing of $p_{T,jet}$ resolution, as shown in Fig.~\ref{fig:jetptresWinset}, into these calculations. This inclusion accounts for the inherent uncertainties associated with the measurement of particle energies in the detector and accommodates effects of fluctuations in heavy-ion events. Model calculations are provided for two regimes, for calculations that
a) include recoiled partons, and b) do not include recoiled partons. When no recoil-tracking is included, the lost jet momentum is removed from the entire system.  This is useful for modeling energy loss in the hard part of the jet.  When recoil-tracking is included, the jets momentum is fully conserved, but this adds both energy and additional background particles to the di-jet. In an experimental analysis, we likely would measure some, but not all of the recoil particles as they are often indistinguishable from background.\
\onecolumngrid\
\begin{figure}[!ht]
	\begin{center}     
		\includegraphics[width=16cm]{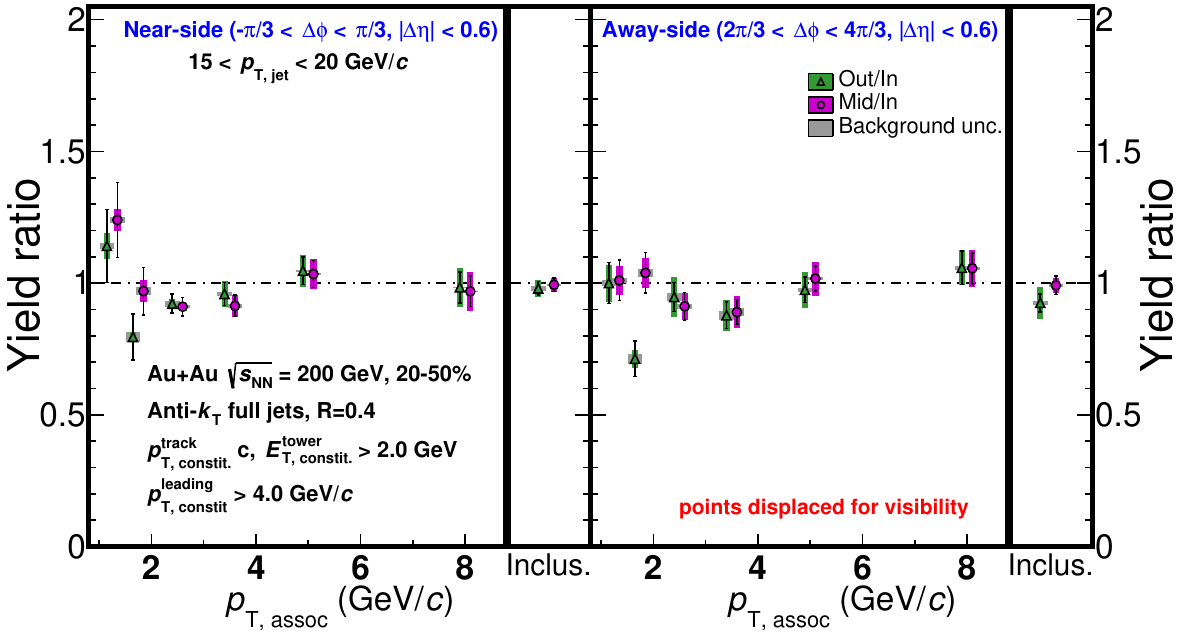}
		\includegraphics[width=16cm]{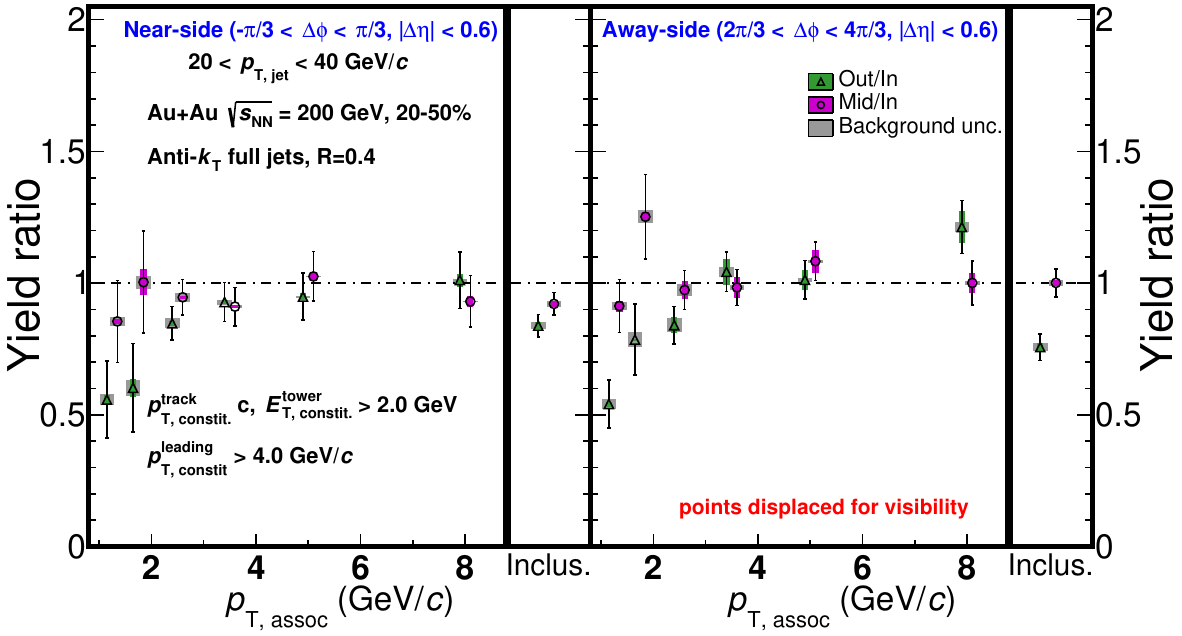}
		\caption{Near-side (left) and away-side (right) associated-yield ratios (of out-of-plane and mid-plane to in-plane) vs \ptAssoc for 15-20 (top) and 20-40 (bottom) \GeV full jets in 20-50\% centrality collisions.  The grey bands describe the systematic uncertainties of the background fits which are non-trivially correlated point-to-point.  The colored bands are scale uncertainties from the JES correction.  Points are displaced for visibility. }
		\label{plot:YieldRatio_C20-50}
	\end{center}
\end{figure}
\twocolumngrid\ 
  Due to the dominant impact of jet-by-jet fluctuations on partonic energy loss over path-length dependence~\cite{Milhano:2015mng,Zapp:2013zya}, JEWEL only predicts a very slight event-plane dependence which is well below the systematic uncertainty in the measurement.  Variations among event-plane orientations were not seen at the 10\% level.  This is therefore consistent with path-length dependence having an insignificant impact compared to jet-by-jet fluctuations in energy loss.  Fluctuations in the density of the medium may also suppress observable path-length dependence and are not included in the JEWEL model.  However,  higher precision JEWEL calculations may be needed to discern any potential event-plane dependent effects.  We thus show the JEWEL comparisons corresponding to the sample integrated over all angles relative to the event plane and compare to the results from data.  Comparisons show that the away-side is well described in terms of both the associated yields and widths by including recoils at low \ptAssoc, while at high \ptAssoc, the yields are better described by not including recoils and the widths have similar results to within uncertainties for both cases.  When looking at the near-side, the widths are quite similar at high \ptAssoc with slightly larger values when including recoils at low \ptAssoc.  For the associated yields at high \ptAssoc, both JEWEL cases are similar to each other, but underestimate the data and including recoils has larger yields that better match the data at low \ptAssoc. 

\label{sect:YieldRatio}
To better quantify and examine the event-plane dependence of the yields, ratios were taken of mid-plane yields relative to in-plane yields and out-of-plane yields relative to in-plane yields.  The advantage of taking ratios is a reduction in systematic uncertainties due to cancellation of uncertainties from several sources.  The propagation of uncertainties was done similarly to that of Sect.~\ref{sect:CorrBkgd}.  The yield ratio is expressed as:

\begin{equation}
 r = \frac{Y_A}{Y_B} = \frac{Y_A^{\rm meas} - Y_A^{\rm bkgd}}{Y_B^{\rm meas} - Y_B^{\rm bkgd}},
 \label{eqn:YieldRatio}
\end{equation}

\noindent where A and B denote different event-plane orientations of the yield.  The statistical errors, coming only from the terms, $Y_A^{\rm meas}$ and $Y_B^{\rm meas}$, which were completely uncorrelated were calculated as: 

\begin{equation}
 \sigma^{\rm stat}_r = \left|r\right|\sqrt{\left(\frac{\sigma_A}{Y_A}\right)^2 + \left(\frac{\sigma_B}{Y_B}\right)^2}.
 \label{eqn:StatRatio}
\end{equation}

\noindent The scale uncertainties, displayed as colored bands on the yield and width plots, are correlated and completely cancel in the ratio.  Uncertainties from the RPF background subtraction are propagated using the covariance matrix from the background fit ($\sigma_{ij}$), but now on Eq.~\ref{eqn:YieldRatio},  which includes correlated background equations in the numerator and denominator of the ratio.  The correlated background uncertainties are given by:

\begin{equation}
  \sigma^{\rm bkgd}_r = \sqrt{\sum_{i=0}^{N} \sum_{j=0}^{N} \frac{\partial r}{\partial p_i} \frac{\partial r}{\partial p_j} \sigma_{ij}}.
  \label{eqn:SysRatio}
\end{equation}

Where, $p_i$'s are the parameters of the RPF fits. Fig.~\ref{plot:YieldRatio_C20-50} shows the near-side (left) and away-side (right) associated-yield ratios of out-of-plane/in-plane and mid-plane/in-plane for 15-20 (top) and 20-40 \GeV (bottom) jets.

For \pttrigrange{15}{20}, the out-of-plane to in-plane associated-yield ratio shows slight enhancement out-of-plane relative to in-plane at low-\ptAssoc, although the effect is small.  This can potentially be due to additional induced gluon radiation that out-of-plane and mid-plane jets would experience relative to in-plane jets, possibly from the longer path length traversed by jets that are not in-plane.   Deviations of the yield ratios from 1.0 are not statistically significant on the away-side, although a small suppression is seen in both mid-plane and out-of plane relative to in-plane on both the near-side and the away-side for \ptassocrange{2.0}{5.0} for out/in.  The suppression is expected to occur at a higher momentum fraction ($z$) of the jet.  On the away-side, the effects are favoring a redistribution of energy from high-momentum constituents to lower momentum constituents.

For \pttrigrange{20}{40} the near-side ratios are consistent with 1.0 with some movement at the two lowest \ptAssoc bins.  On the away-side, mid/in is consistent with 1.0 with a little enhancement and out/in has an enhancement at high-\ptAssoc and suppression at low-\ptAssoc. This observation contradicts the expectations.  If in-plane jets interact less, we expect ratios to be $< 1.0$ at high-\ptAssoc and $> 1.0$ at low-\ptAssoc.  This is a reminder of the competing effects in the analyzed momentum range, and it is an indication that the expected path length effects due to jet energy loss are dominated by the fluctuations in the medium.

Alternatively, the initial geometry within specific centrality bins may be fluctuating at a larger magnitude than any possible event-plane dependence~\cite{Noronha-Hostler-RAAv2,PhysRevLett.116.252301}.\  The possibility of this occurrence can be studied by looking at the initial configuration and selecting low and high ellipticity events by  using the $Q_n$ flow vector found within a selected centrality range.

To study the impact of surface bias and event-plane resolution, a check was performed to investigate the systematic change in the ratio of yields (out/in, mid/in) with the angle of the event plane by fitting a constant to Fig.~\ref{plot:YieldRatio_C20-50} for jets with \pttrigrange{15}{20} (top) and \pttrigrange{20}{40} (bottom).  The systematic uncertainties are treated as uncorrelated point-to-point and added to the statistical uncertainties in quadrature.  The results are shown in Tabs.~\ref{Tab:RatioFits_J15-20} and~\ref{Tab:RatioFits_J20-40} and are consistent with one.  Care should be taken when interpreting the results, as various effects can be in play and medium modifications could give way to a \ptAssoc dependence. This \ptAssoc dependence can be seen as an enhancement of associated yields for $\ptAssoc \geq$ 2 \GeV on the near-side from the geometric kinematic biases due to the ``hard-core" requirement of jet reconstruction.\   
\onecolumngrid\
\begin{table}[!htbp]
\caption{Results of fits to Fig.\ref{plot:YieldRatio_C20-50} (top panel: 15-20 \GeV jets) to a constant $c$, the $\chi^2$ over the number of degrees of freedom (NDF), the number of standard deviations $\sigma$ of $c$ from one, and the range of $c$ within a 90\% confidence limit (CL).}
\label{Tab:RatioFits_J15-20}
\centering 
\begin{tabular}{c | c c | c c} 
\hline
& \multicolumn{2}{c |}{Near-side}  & \multicolumn{2}{c}{Away-side}  \\
parameter & $Y_{\mathrm{out}} / Y_{\mathrm{in}}$  & $Y_{\mathrm{mid}} / Y_{\mathrm{in}}$   & $Y_{\mathrm{out}} / Y_{\mathrm{in}}$  & $Y_{\mathrm{mid}} / Y_{\mathrm{in}}$  \\ \hline

$c$          & 0.93 $\pm$ 0.042 & 0.949 $\pm$ 0.038 & 0.89 $\pm$ 0.065 & 0.997 $\pm$ 0.066 \\
$\chi^2$/NDF & 0.68             & 0.67              & 0.71             & 0.18 \\
$\sigma$     & -1.6             & -1.3              & -1.7             & -0.1 \\
90\% CL      & 0.86 -- 1.00     & 0.89 -- 1.01      & 0.78 -- 1.00     & 0.94 -- 1.05 \\
\hline
\end{tabular}
\end{table}

\begin{table}[!htbp]
\caption{Results of fits to Fig.\ref{plot:YieldRatio_C20-50} (bottom panel: 20-40 \GeV jets) to a constant $c$, the $\chi^2$ over the number of degrees of freedom (NDF), the number of standard deviations $\sigma$ of $c$ from one, and the range of $c$ within a 90\% confidence limit (CL).}
\label{Tab:RatioFits_J20-40}
\centering 
\begin{tabular}{c | c c | c c} 
\hline
& \multicolumn{2}{c |}{Near-side}  & \multicolumn{2}{c}{Away-side}  \\
parameter & $Y_{\mathrm{out}} / Y_{\mathrm{in}}$  & $Y_{\mathrm{mid}} / Y_{\mathrm{in}}$   & $Y_{\mathrm{out}} / Y_{\mathrm{in}}$  & $Y_{\mathrm{mid}} / Y_{\mathrm{in}}$  \\ \hline

$c$          & 0.874 $\pm$ 0.043 & 0.937 $\pm$ 0.043 & 0.752 $\pm$ 0.064 & 1.02 $\pm$ 0.075 \\
$\chi^2$/NDF & 1.4               & 0.19              & 2.1               & 0.49 \\
$\sigma$     & -2.9              & -1.5              & -3.9              & 0.2 \\
90\% CL      & 0.77 -- 0.98      & 0.90 -- 0.97      & 0.56 -- 0.94      & 0.91 -- 1.12 \\
\hline
\end{tabular}
\end{table}
\twocolumngrid\
\section{Conclusions} \label{sect:Summary}
The measurement of jet-hadron correlations relative to the event plane is reported for the 20-50\% most central events in \AuAu collisions at \sNN = 200 GeV in STAR.\ Partonic interactions are directly related to the distance traversed in the medium, so it is expected that medium-induced jet modifications should depend on the path length. The angle of the jet, measured with respect to the event plane, is correlated on average with the jet's path length through the medium. In this analysis, the average path length of away-side jets is potentially increased due to the surface bias of the near-side trigger jet. This work utilizes the RPF background-subtraction method to remove the event-plane dependent background while reducing uncertainties and assumptions associated with previous background-subtraction techniques. Associated yields, their ratios, and jet-peak widths are extracted for each event-plane orientation and compared with different average path lengths and JEWEL model calculations. JEWEL performs better in describing the associated yields and widths at higher \ptAssoc when recoil partons are not included. Conversely, including recoil partons leads to JEWEL providing a better description of the lower \ptAssoc region. This study highlights the importance of conducting further tuning of Monte Carlo simulations to accurately describe the results in this analysis for jets that are biased towards hard-fragmented jets due to the HT-trigger and hard-core constituent requirements.

Within the precision of the current measurement at \sNN = 200 GeV, the associated yields and jet-peak widths show no dependence on the event plane. The ratios derived from the associated yields are used to quantify the differences, but they do not deviate significantly from 1.0. For the \pttrigrange{20}{40}, there were indications of potential modifications observed in the inclusive bin of \ptassocrange{1.0}{10}.
The results presented in this study align with the findings observed in hadron-hadron and jet-hadron correlations reported in
Ref.~\cite{Nattrass:2016cln} for RHIC and  Ref.~\cite{PhysRevC.101.064901} for LHC energies.  The lack of clear event-plane dependence in our data indicates that any dependence of these modifications on the average path-length is less than our experimental uncertainties.  

\section{Acknowledgements} \label{sect:Acknowledgements}

We thank the RHIC Operations Group and RCF at BNL, the NERSC Center at LBNL, and the Open Science Grid consortium for providing resources and support.  This work was supported in part by the Office of Nuclear Physics within the U.S. DOE Office of Science, the U.S. National Science Foundation, National Natural Science Foundation of China, Chinese Academy of Science, the Ministry of Science and Technology of China and the Chinese Ministry of Education, the Higher Education Sprout Project by Ministry of Education at NCKU, the National Research Foundation of Korea, Czech Science Foundation and Ministry of Education, Youth and Sports of the Czech Republic, Hungarian National Research, Development and Innovation Office, New National Excellency Programme of the Hungarian Ministry of Human Capacities, Department of Atomic Energy and Department of Science and Technology of the Government of India, the National Science Centre and WUT ID-UB of Poland, the Ministry of Science, Education and Sports of the Republic of Croatia, German Bundesministerium f\"ur Bildung, Wissenschaft, Forschung and Technologie (BMBF), Helmholtz Association, Ministry of Education, Culture, Sports, Science, and Technology (MEXT), Japan Society for the Promotion of Science (JSPS) and Agencia Nacional de Investigaci\'on y Desarrollo (ANID) of Chile.

\bibliographystyle{apsrev4-2}

\begin{thebibliography}{10}
  \bibitem{Adcox:2004mh}
  K.~Adcox {\em et~al.}, ``{Formation of dense partonic matter in relativistic nucleus-nucleus collisions at RHIC: Experimental evaluation by the PHENIX collaboration},'' {\em Nucl. Phys. A}, vol.~757, pp.~184--283, 2005.
  \bibitem{Adams:2005dq}
  J.~Adams {\em et~al.}, ``{Experimental and theoretical challenges in the search for the quark gluon plasma: The STAR collaboration's critical assessment of the evidence from RHIC collisions},'' {\em Nucl. Phys.}, vol.~A757, pp.~102--183, 2005.
  \bibitem{Arsene:2004fa}
  I.~Arsene {\em et~al.}, ``{Quark Gluon Plasma an Color Glass Condensate at RHIC? The perspective from the BRAHMS experiment},'' {\em Nucl. Phys.}, vol.~A757, pp.~1--27, 2005.
  \bibitem{Back:2004je}
  B.~B. Back {\em et~al.}, ``{The PHOBOS perspective on discoveries at RHIC},'' {\em Nucl. Phys.}, vol.~A757, pp.~28--101, 2005.
  \bibitem{Shuryak:2004cy}
  E.~V. Shuryak, ``{What RHIC experiments and theory tell us about properties of quark-gluon plasma?},'' {\em Nucl. Phys. A}, vol.~750, pp.~64--83, 2005.
  \bibitem{Lee:2005gw}
  T.~D. Lee, ``{The strongly interacting quark-gluon plasma and future physics},'' {\em Nucl. Phys. A}, vol.~750, pp.~1--8, 2005.
  \bibitem{Gyulassy:2004zy}
  M.~Gyulassy and L.~McLerran, ``{New forms of QCD matter discovered at RHIC},'' {\em Nucl. Phys. A}, vol.~750, pp.~30--63, 2005.
  \bibitem{Heinz:2005zg}
  U.~W. Heinz, ``{'RHIC serves the perfect fluid': Hydrodynamic flow of the QGP},'' in {\em {Workshop on Extreme QCD}}, 12 2005.
  \bibitem{Blau:2005pk}
  S.~K. Blau, ``{A string-theory calculation of viscosity could have surprising applications},'' {\em Phys. Today}, vol.~58N5, pp.~23--24, 2005.
  \bibitem{Riordan:2006df}
  M.~Riordan and W.~A. Zajc, ``{The first few microseconds},'' {\em Sci. Am.}, vol.~294N5, pp.~24--31, 2006.
  \bibitem{Gyulassy:1990bh}
  M.~Gyulassy and M.~Plumer, ``{Jet quenching as a probe of dense matter},'' {\em Nucl. Phys. A}, vol.~527, pp.~641--644, 1991.
  \bibitem{Gyulassy:2003mc}
  M.~Gyulassy {\em et~al.}, ``{Jet quenching and radiative energy loss in dense nuclear matter},'' {\em Quark–Gluon Plasma 3}, p.~123–191, Jan 2004.
  \bibitem{Wang:1991xy}
  X.N.~Wang and M.~Gyulassy, ``{Gluon shadowing and jet quenching in A + A collisions at $\sqrt{s_{\rm NN}} = 200$ GeV },'' {\em Phys. Rev. Lett.}, vol.~68, pp.~1480--1483, 1992.
  \bibitem{Adams:2003im}
  J.~Adams {\em et~al.}, ``{Evidence from d + Au measurements for final state suppression of high $p_T$ hadrons in Au+Au collisions at RHIC },'' {\em Phys. Rev. Lett.}, vol.~91, p.~072304, 2003.
  \bibitem{Adams:2006yt}
  J.~Adams {\em et~al.}, ``{Direct observation of dijets in central Au+Au collisions at $\sqrt{s_{\rm NN}} = 200$ GeV},'' {\em Phys. Rev. Lett.}, vol.~97, p.~162301, 2006.
  \bibitem{Adler:2006hu}
  S.~S. Adler {\em et~al.}, ``{Common suppression pattern of eta and pi0 mesons at high transverse momentum in Au+Au collisions at $\sqrt{s_{\rm NN}} = 200$ GeV},'' {\em Phys. Rev. Lett.}, vol.~96, p.~202301, 2006.
  \bibitem{Chatrchyan:2011sx}
  S.~Chatrchyan {\em et~al.}, ``{Observation and studies of jet quenching in PbPb collisions at nucleon-nucleon center-of-mass energy = 2.76 TeV},'' {\em Phys. Rev.}, vol.~C84, p.~024906, 2011.
  \bibitem{Chatrchyan:2012gt}
  S.~Chatrchyan {\em et~al.}, ``{Studies of jet quenching using isolated-photon+jet correlations in PbPb and $pp$ collisions at $\sqrt{s_{NN}}=2.76$ TeV},'' {\em Phys. Lett. B}, vol.~718, pp.~773--794, 2013.
  \bibitem{CMS:2016uxf}
  V.~Khachatryan {\em et~al.}, ``{Measurement of inclusive jet cross sections in $pp$ and PbPb collisions at $\sqrt{s_{NN}}=$ 2.76 TeV},'' {\em Phys. Rev. C}, vol.~96, no.~1, p.~015202, 2017.
  \bibitem{Chatrchyan:2012gw}
  S.~Chatrchyan {\em et~al.}, ``{Measurement of jet fragmentation into charged particles in $pp$ and PbPb collisions at $\sqrt{s_{NN}}=2.76$ TeV},'' {\em JHEP}, vol.~10, p.~087, 2012.
  \bibitem{CMS:2021vui}
  A.~M. Sirunyan {\em et~al.}, ``{First measurement of large area jet transverse momentum spectra in heavy-ion collisions},'' {\em JHEP}, vol.~05, p.~284, 2021.
  \bibitem{PhysRevLett.105.252303}
  G. Aad {\em et~al.}, ``{Observation of a Centrality-Dependent Dijet Asymmetry in Pb-Pb Collisions at $\sqrt{{s}_{\mathrm{NN}}}=2.76$ TeV with the ATLAS Detector at the LHC},'' \em{Phys. Rev. Lett.}, vol.~105, p.~252303, 2010.
\bibitem{JHEP03.2014.13}
  The ALICE collaboration, ``{Measurement of charged jet suppression in Pb-Pb collisions at $\sqrt{{s}_{\mathrm{NN}}}=2.76$ TeV}'',
  \em{J. High Energ. Phys.}, 2014, 13.
  \bibitem{Wiedemann:2009sh}
  U.~A. Wiedemann, ``{Jet Quenching in Heavy Ion Collisions},'' {\em Landolt-Börnstein - Group I Elementary Particles, Nuclei and Atoms}, pp.~521--562, 2010.
  \newblock [Landolt-Bornstein23,521(2010)].
  \bibitem{doi:10.1146/annurev.nucl.50.1.37}
  R.~Baier, D.~Schiff, and B.~G. Zakharov, ``Energy loss in perturbative qcd,'' {\em Annual Review of Nuclear and Particle Science}, vol.~50, no.~1, pp.~37--69, 2000.
  \bibitem{Khachatryan:2016erx}
  V.~Khachatryan {\em et~al.}, ``{Correlations between jets and charged particles in PbPb and pp collisions at $ \sqrt{s_{\mathrm{NN}}}=2.76 $ TeV},'' {\em JHEP}, vol.~02, p.~156, 2016.
  \bibitem{CMS:2021nhn}
  A.~M. Sirunyan {\em et~al.}, ``{In-medium modification of dijets in PbPb collisions at $ \sqrt{s_{\mathrm{NN}}} $ = 5.02 TeV},'' {\em JHEP}, vol.~05, p.~116, 2021.
  \bibitem{Sirunyan_2018}
  A.~M. Sirunyan {\em et~al.}, ``Measurement of the splitting function in Pb-Pb collisions at 5.02 TeV,'' {\em Physical Review Letters}, vol.~120, Apr 2018.
  \bibitem{ALargeIonColliderExperiment:2021mqf}
  S.~Acharya \textit{et al.},
  ``Measurement of the groomed jet radius and momentum splitting fraction in pp and Pb$-$Pb collisions at $\sqrt{s_{NN}} = 5.02$ TeV,''
  Phys. Rev. Lett. 128, no.10, 102001 (2022)
  doi:10.1103/PhysRevLett.128.102001
  [arXiv:2107.12984 [nucl-ex]].
  \bibitem{ATLASOpenAng}
  G. Aad et al. (ATLAS Collaboration),
  ``Measurement of substructure-dependent jet suppression in Pb+Pb collisions at 5.02 TeV with the ATLAS detector''
  Phys. Rev. C 107, 054909 
  \bibitem{STAR:2021kjt}
  M.~S. Abdallah {\em et~al.}, ``{Differential measurements of jet substructure and partonic energy loss in Au+Au collisions at $\sqrt {S_{NN}}$ =200 GeV},'' 9 2021.
  \bibitem{Liou:2013qya}
  T.~Liou, A.~H. Mueller, and B.~Wu, ``{Radiative $p_\bot$-broadening of high-energy quarks and gluons in QCD matter},'' {\em Nucl. Phys. A}, vol.~916, pp.~102--125, 2013.
  \bibitem{PhysRevC.76.034904}
  S.~S. Adler {\em et~al.}, ``Detailed study of high ${p}_{T}$ neutral pion suppression and azimuthal anisotropy in $\mathrm{Au}+\mathrm{Au}$ collisions at $\sqrt{s_{\mathit{NN}}} = 200$ gev,'' {\em Phys. Rev. C}, vol.~76, p.~034904, Sep 2007.
  \bibitem{PhysRevC.101.064901}
  S.~Acharya {\em et~al.}, ``Jet-hadron correlations measured relative to the second order event plane in pb-pb collisions at $\sqrt{{s}_{\rm NN}}=2.76\phantom{\rule{0.16em}{0ex}}\mathrm{TeV}$,'' {\em Phys. Rev. C}, vol.~101, p.~064901, Jun 2020.
  \bibitem{Cacciari:2008gp}
  M.~Cacciari, G.~P. Salam, and G.~Soyez, ``{The Anti-k(t) jet clustering algorithm},'' {\em JHEP}, vol.~04, p.~063, 2008.
  \bibitem{Sharma:2015qra}
  N.~Sharma, J.~Mazer, M.~Stuart, C.~Nattrass, ``{Background subtraction methods for precision measurements of di-hadron and jet-hadron correlations in heavy ion collisions},'' {\em Phys. Rev.}, vol.~C93, no.~4, p.~044915, 2016.
  \bibitem{Zapp:2013vla}
  K.~C. Zapp, ``{JEWEL 2.0.0: directions for use},'' {\em Eur. Phys. J.}, vol.~C74, no.~2, p.~2762, 2014.
  
  \bibitem{Ackermann:2002ad}
  K.~H. Ackermann {\em et~al.}, ``{STAR detector overview},'' {\em Nucl. Instrum. Meth.}, vol.~A499, pp.~624--632, 2003.
  
  \bibitem{Anderson:2003ur}
  M.~Anderson {\em et~al.}, ``{The Star time projection chamber: A Unique tool for studying high multiplicity events at RHIC},'' {\em Nucl. Instrum. Meth.}, vol.~A499, pp.~659--678, 2003.
  
  \bibitem{BEDDO2003725}
  M.~Beddo {\em et~al.}, ``The star barrel electromagnetic calorimeter,'' {\em Nuclear Instruments and Methods in Physics Research Section A: Accelerators, Spectrometers, Detectors and Associated Equipment}, vol.~499, no.~2, pp.~725 -- 739, 2003.
  
  \bibitem{Abelev:2013fn}
  B.~Abelev {\em et~al.}, ``{Measurement of the inclusive differential jet cross section in $pp$ collisions at $\sqrt{s} = 2.76$ TeV},'' {\em Phys. Lett.}, vol.~B722, pp.~262--272, 2013.
  
  \bibitem{PhysRevLett.119.062301}
  L.~Adamczyk {\em et~al.}, ``Dijet imbalance measurements in $\mathrm{Au}+\mathrm{Au}$ and $pp$ collisions at $\sqrt{{s}_{\rm NN}}=200\text{ }\mathrm{GeV}$ at star,'' {\em Phys. Rev. Lett.}, vol.~119, p.~062301, Aug 2017.
  
  \bibitem{PhysRevLett.115.092002}
  L.~Adamczyk {\em et~al.}, ``Precision measurement of the longitudinal double-spin asymmetry for inclusive jet production in polarized proton collisions at $\sqrt{s}=200\text{ }\mathrm{GeV}$,'' {\em Phys. Rev. Lett.}, vol.~115, p.~092002, Aug 2015.
  
  \bibitem{Beddo:2002zx}
  M.~Beddo {\em et~al.}, ``{The STAR Barrel Electromagnetic Calorimeter},'' {\em Nucl. Instrum. Meth.}, vol.~A499, pp.~725--739, 2003.
  
  \bibitem{PhysRevC.96.024905}
  L.~Adamczyk {\em et~al.}, ``Measurements of jet quenching with semi-inclusive hadron+jet distributions in $\text{Au}+\text{Au}$ collisions at $\sqrt{{s}_{NN}}=200$ gev,'' {\em Phys. Rev. C}, vol.~96, p.~024905, Aug 2017.
  
  \bibitem{Poskanzer:1998yz}
  A.~M. Poskanzer and S.A.~Voloshin, ``{Methods for analyzing anisotropic flow in relativistic nuclear collisions},'' {\em Phys.Rev.}, vol.~C58, pp.~1671--1678, 1998.
  
  \bibitem{CASTILHO2018369}
  W.~M. Castilho, W.~Qian, Y.~Hama, and T.~Kodama, ``Event-plane dependent di-hadron correlations with harmonic $v_n$ subtraction in a hydrodynamic model,'' {\em Physics Letters B}, vol.~777, pp.~369--373, 2018.
  
  \bibitem{Agakishiev:2014ada}
  H.~Agakishiev {\em et~al.}, ``{Event-plane-dependent dihadron correlations with harmonic $v_n$ subtraction in Au + Au collisions at $\sqrt{s_{NN}}=200$ GeV},'' {\em Phys. Rev.}, vol.~C89, no.~4, p.~041901, 2014.
  
  \bibitem{Adams_2005}
  J.~Adams {\em et~al.}, ``Azimuthal anisotropy in au+au collisions at $\sqrt{s_{NN}}=200$ gev,'' {\em Physical Review C}, vol.~72, p.~014904, Jul 2005.
  
  \bibitem{Barrette:1997pt}
  J.~Barrette {\em et~al.}, ``{Proton and pion production relative to the reaction plane in Au + Au collisions at AGS energies},'' {\em Phys. Rev.}, vol.~C56, pp.~3254--3264, 1997.
  
  \bibitem{PhysRevC.55.1420}
  J.~Barrette {\em et~al.}, ``Energy and charged particle flow in $10.8a$ gev/$c$ au+au collisions,'' {\em Phys. Rev. C}, vol.~55, pp.~1420--1430, Mar 1997.
  
  \bibitem{Voloshin:2008dg}
  S.~A. Voloshin, A.~M. Poskanzer, and R.~Snellings, ``{Collective phenomena in non-central nuclear collisions},'' {\em Landolt-Bornstein}, vol.~23, pp.~293--333, 2010.
  
  \bibitem{Wang:2012bga}
  Q.~Wang, {\em {Charge Multiplicity Asymmetry Correlation Study Searching for Local Parity Violation at RHIC for STAR collaboration}}.
  \newblock PhD thesis, Kansas U., New York, 2012.
  
  \bibitem{Bielcikova:2003ku}
  J.~Bielcikova {\em et~al.}, ``{Elliptic flow contribution to two particle correlations at different orientations to the reaction plane},'' {\em Phys.Rev.}, vol.~C69, p.~021901(R), 2004.
  
  \bibitem{Voloshin:1994mz}
  S.~Voloshin and Y.~Zhang, ``{Flow study in relativistic nuclear collisions by Fourier expansion of Azimuthal particle distributions},'' {\em Z.Phys.}, vol.~C70, pp.~665--672, 1996.
  
  \bibitem{Abbas:2013taa}
  E.~Abbas {\em et~al.}, ``{Performance of the ALICE VZERO system},'' {\em JINST}, vol.~8, p.~P10016, 2013.
  
  \bibitem{Barrette:1996rs}
  J.~Barrette {\em et~al.}, ``{Energy and charged particle flow in a 10.8-AGeV/c Au+Au collisions},'' {\em Phys.Rev.}, vol.~C55, pp.~1420--1430, 1997.
  
  \bibitem{Cacciari:2011ma}
  M.~Cacciari, G.~P. Salam, and G.~Soyez, ``{FastJet User Manual},'' {\em Eur.Phys.J.}, vol.~C72, p.~1896, 2012.
  
  \bibitem{Cacciari:2008gn}
  M.~Cacciari, G.~P. Salam, and G.~Soyez, ``{The Catchment Area of Jets},'' {\em JHEP}, vol.~04, p.~005, 2008.
  
  \bibitem{Aad:2014fla}
  G.~Aad {\em et~al.}, ``{Measurement of event-plane correlations in $\sqrt{s_{NN}}=2.76$ TeV lead-lead collisions with the ATLAS detector},'' {\em Phys. Rev.}, vol.~C90, no.~2, p.~024905, 2014.
  
  \bibitem{Nattrass_2018}
  C.~Nattrass and T.~Todoroki, ``Event plane dependence of the flow modulated background in dihadron and jet-hadron correlations in heavy ion collisions,'' {\em Physical Review C}, vol.~97, May 2018.
  
  \bibitem{Nattrass_Reexam2018}
  C.~Nattrass, ``Re-examining the iconic dihadron correlation measurement demonstrating jet quenching,'' {\em Physical Review C}, vol.~97, Mar 2018.
  
  \bibitem{Nattrass:2016cln}
  C.~Nattrass, N.~Sharma, J.~Mazer, M.~Stuart, A.~Bejnood, ``{Disappearance of the Mach Cone in heavy ion collisions},'' {\em Phys. Rev.}, vol.~C94, no.~1, p.~011901(R), 2016.
  
  \bibitem{Zapp_2009}
  K.~Zapp, G.~Ingelman, J.~Rathsman, J.~Stachel, and U.~A. Wiedemann, ``A monte carlo model for `jet quenching','' {\em The European Physical Journal C}, vol.~60, mar 2009.
  
  \bibitem{Milhano:2015mng}
  J.~G. Milhano and K.~C. Zapp, ``{Origins of the di-jet asymmetry in heavy ion collisions},'' {\em Eur. Phys. J.}, vol.~C76, no.~5, p.~288, 2016.
  
  \bibitem{Zapp:2013zya}
  K.~C. Zapp, ``{Geometrical aspects of jet quenching in JEWEL},'' {\em Phys.Lett.}, vol.~B735, pp.~157--163, 2014.
  
  \bibitem{Noronha-Hostler-RAAv2}
  J.~Noronha-Hostler, ``Solving the ${R}_{AA}\ensuremath{\bigotimes}v_2$ puzzle,'' {\em Journal of Physics: Conference Series}, vol.~736, Aug. 2016.
  
  \bibitem{PhysRevLett.116.252301}
  J.Noronha-Hostler, B.Betz, J.Noronha, M.Gyulassy,``Event-by-event $\mathrm{Hydrodynamics}+\mathrm{Jet}$ energy loss: A solution to the ${R}_{AA}\bigotimes{v}_{2}$ puzzle,'',
  {\em Phys. Rev. Lett.}, vol.~116, p.~252301, Jun 2016.

  \end{thebibliography}

\end{document}